\documentclass[article,twocolumn,amsmath,floats,showpacs,nofootinbib]{revtex4}
\usepackage[usenames]{color}
\usepackage{graphicx}
\usepackage{epstopdf}
\usepackage{textcomp}
\usepackage{hyperref}
\usepackage{multirow}
\usepackage{tikz}
\usepackage{rotating}
\hypersetup{colorlinks=true}

\begin{document}

\title{Point-contact studies of semiconductor $PbTe-PbS$ superconducting superlattice as a model of HTS}

\author{N. L. Bobrov, L. F. Rybal'chenko,  V. V. Fisun, I. K. Yanson, O. A. Mironov, S. V. Chistyakov, V. V. Zorchenko, A. Yu. Sipatov, and A. I. Fedorenko}
\affiliation{B.I.~Verkin Institute for Low Temperature Physics and
Engineering, of the National Academy of Sciences
of Ukraine, prospekt Lenina, 47, Kharkov 61103, Ukraine, Institute of Radiophysics and Electronics, Academy of Sciences of the Ukrainian SSR, Kharkov and V. I. Lenin Polytechnical Institute, Kharkov
Email address: bobrov@ilt.kharkov.ua}
\published {(\href{http://fntr.ilt.kharkov.ua/fnt/pdf/16/16-12/f16-1531r.pdf}{Fiz. Nizk. Temp.}, \textbf{16}, 1531 (1990)); (Sov. J. Low Temp. Phys., \textbf{16}, 862 (1990)}
\date{\today}

\begin{abstract}Superconducting (SC) superlattices (SL) of the epitaxial semiconductor $PbTe-PbS$/(001) $KCl$ with two-dimensional ordered misfit dislocation (MD) networks across heteroboundaries (HB) are studied, and the obtained results are compared with the corresponding data for $YBaCuO$. An analysis of the fluctuational region showed that the pairing of electrons in SL takes place at the sites of dislocation networks. In the region of SC fluctuations, the IVC derivatives of point-contacts (PC) based on SL and HTS display a metal-insulator transition associated with delocalization of carriers from the vicinity of the structural elements responsible for the emergence of SC. The critical current density in a superlattice is determined, and a mechanism is proposed for formation of excess current on the IVC of PC based on SL and HTS. The values of the energy gap $\Delta_0$ at low temperatures in units of $kT_c$ are found to be anomalously large for SL and HTS. For anisotropic three-dimensional SL with a strong bond between two-dimensional SC planes, two gaps emerge from the MD. For quasi-two-dimensional SL with a weak bond between two-dimensional SC planes, the $\Delta(T)$ dependence is nonmonotonic, and the fluctuation region for these SL and for HTS reveals a gap that is practically independent of temperature. The IVC derivatives of a $Cu$-SL PC reveal an oscillating structure which is restricted to an energy of $90\ meV$ and is apparently associated with the electron-phonon interaction. It is assumed that there is a similarity between the SC mechanisms in SL and HTS.

\end{abstract}

\maketitle

\section{INTRODUCTION}
Following the discovery of oxide high-temperature superconductors (HTS) with a typical layered structure, interest has grown considerably in recent years towards the properties of superconducting superlattices (SL), whose investigation may turn out to be quite useful for understanding the physics of HTS \cite{1, 2, 3, 4, 5}. The present paper is devoted to the investigation of properties of the superconducting SL of the semiconductor $PbTe-PbS$/(001) KC1 with the help of point contacts (PC). It will be shown below that these SL and HTS have a certain resemblance in the elements of their crystal structure, as well as in the peculiarities of their PC spectra \cite{3} and the temperature dependences of the critical currents for films \cite{4}. This makes it possible to consider the SL of lead chalcogenides as a physical analog of HTS.

Lead chalcogenides have an $NaCl$ type cubic lattice and their epitaxial films have a fairly high degree of technological perfection. Lead atoms are located at the centers of octahedra formed by chalcogen atoms. In a similar manner, copper atoms in HTS are surrounded by tetrahedra formed by oxygen atoms. Lead chalcogenides are characterized by a strong non-stoichiometry (excess or deficit of chalcogens) determining the concentration and mobility of charge carriers. To a certain extent, this is equivalent to the role of oxygen deficit in HTS.

The layered structure of HTS can easily be modeled with the help of SL from thin layers of two different lead chalcogenides. Through an appropriate choice of the lead chalcogenide pairs and a variation of the thickness of their layers in SL, it is possible to produce uniform elastic pseudomorphic deformations of films over their thickness, as well as nonuniform deformations associated with misfit dislocations (MD) at the interfaces between layers. It was shown earlier in Ref. \cite{2} that the superconductivity (SC) of such compositions is determined by regular square networks of MD.

\begin{figure}[]
\includegraphics[width=8cm,angle=0]{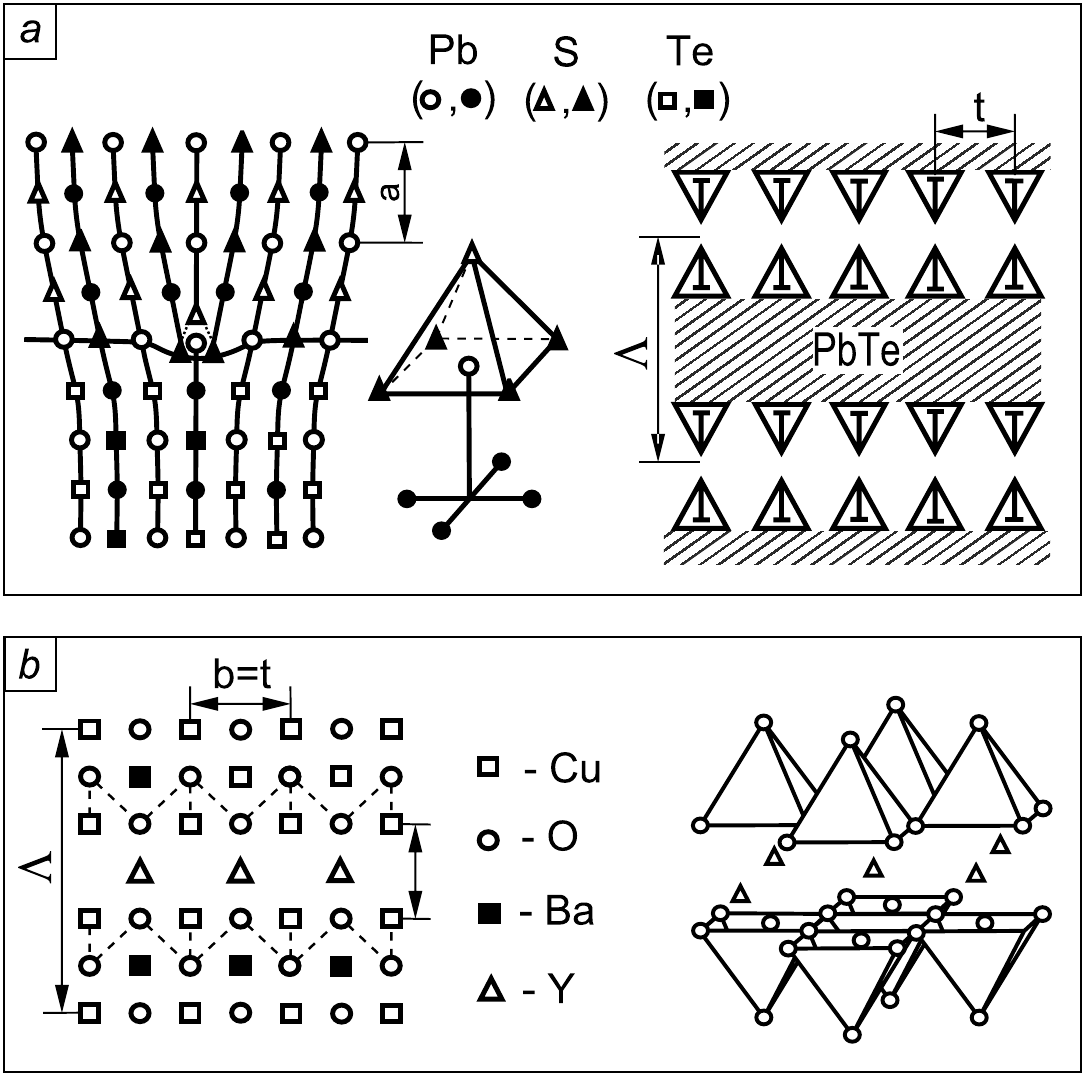}
\caption[]{(\emph{a}) Schematic representation of the nucleus of misfit dislocations and packing of layers in the SL $PbTe-PbS$, (\emph{b}) structural elements of a $YBaCuO$ type HTS. The circles, triangles, and squares on the left-hand side show the projections of the corresponding atomic chains. Figure 1a also shows the local configuration of atoms at a dislocation site marked by triangles on the right-hand side. The copper-oxygen tetrahedra are marked by dashed lines on the left-hand side of Fig.~1\emph{b}. The size \emph{a} in die upper part of the figure derates the lattice parameter of the $PbS$ lattice.}
\label{Fig1}
\end{figure}

It has been practically established that in systems of the type 1-2-3 and in $Bi$ and $Tl$ families, the high critical temperatures $T_c$ are associated with twinned $CuO_2$ planes \cite{6,7} separated by a layer of metal atoms (e.g., $Y$ in the system $YBaCuO$ and $Sr$, $Ca$ in the system $Bi(Sr, Ca)CuO)$. The similarity of these elements of perovskite structure is responsible for the closeness of the superconducting transition temperatures of various oxides \cite{6}. It has been reported in Ref. \cite{7} that the introduction of additional $CuO_2$ planes in the system $TlCaBaCuO$ leads to an increase in $T_c$ to $162\ K$ (zero resistance). The $CuO_2$ layers form tetrahedra with the adjoining oxygen layers. These tetrahedra have copper atoms at their base and are arranged in a staggered manner in a plane parallel to the basal plane $a-b$ (see Fig.\ref{Fig1}$b$). In a certain sense, the structure of the SL $PbTe-PbS$ is similar to the HTS structure. Figure \ref{Fig1}$a$ shows schematically the proposed arrangement of atoms in the vicinity of the MD line. Apparently, the role of tetrahedra is played here by dislocation sites at which lead atoms are surrounded by tetrahedra formed by chalcogen atoms. In the family $LaSrCuO$ also, superconductivity is associated with $CuO_2$ planes, but the packing of atoms differs considerably from that in other HTS compounds, and the pattern of onset of SC is less clear than in 1-2-3 type compounds. Moreover, there is no direct analogy between the $LaSrCuO$ structure and SL. Hence in the following we shall compare SL only with 1-2-3 type materials.

Since the work function of $PbS$ is much lower (by 0.6~$eV$) \cite{8} than that of $PbTe$, electrons from $PbS$ layers must go over to $PbTe$ layer, leading to alternating electron ($PbTe$) and hole ($PbS$) layers. Of course, we must take into account the fact that the work functions in thin films may differ significantly from their values for bulk materials. However, in view of the closeness of properties of $PbTe$ and $PbS$, a significant variation or a sign reversal of the difference in the work functions of $PbTe$ and $PbS$ can be ruled out in multilayer compositions. In view of the electron nature of the conductivity of the entire SL, we can obviously treat $PbS$ as in "insulating" interlayer (from the point of view of SC). The introduction of the MD network must lead to a considerable variation of the energy spectrum of charge carriers in a layered composite. The emergence of a contravariant band structure of the initial semiconductors also cannot be rule out \cite{9}. Nonuniform MD deformation fields lead to a modulation of the valence and conduction band edges. The estimate of the potential were depth for electrons near the dislocation network has the same order of magnitude as the Fermi energy in SL, hence one can expect the emergence of electron states localized at the MD network (dislocation band), which facilitates the emergence of SC in the $PbTe-PbS$.

It was shown in Ref. \cite{9} that in the vicinity of MD, a network of conducting electron states of reduced dimensionality is formed in SL. These states are specific boundary states existing in the conduction band of the initial layers only in a limited interval of energies and transverse wave vectors. The density of states on the dislocation network can attain large values, which is quite significant for SC in the model of the resonance scattering observed in $p-PbTe$ $\langle Tl \rangle$ \cite{10}.

The mechanism of superconductivity is not clear at present. It may be due to the interaction of a narrow dislocation band with the bulk states. Such arguments were put forth for HTS materials \cite{11,12}. The fact that the two-layer sandwich $PbS-PbTe$ does not have superconductivity, at least for $T>1.5\ K$, is of fundamental importance. The introduction of a third $PbS$ layer results in the emergence of SC with $T_c>(3.5-5.5)\ K$ \cite{4}. This means that the interaction of at least two heteroboundaries stabilizes the SC induced in them. Such an assumption was confirmed by the preliminary investigations of temperature and angular dependences of the upper critical fields $H_{c2}$ \cite{4,5} which show that for layers of thickness larger than the period of the dislocation structure, the SC is determined by the interaction of adjacent dislocation networks across the $PbTe$ layer. For small thicknesses, the $PbTe$ layer with two MD networks can be treated as a single SC quasi- two-dimensional layer, and $T_c$ depends on the interaction of such layers through the $PbS$ interlayers. The maximum values of $T_c$ are attained (for a given stoichiometry of layers and nearly equal thicknesses of $PbTe$ and $PbS$) in samples with $t_{PbTe}=170\ {\AA}$, $t_{PbS}=180\ {\AA}$. It is this type of samples (with a superstructure period $\Lambda=t_{PbTe}+t_{PbS}=350\ {\AA}$) that were used in the investigations whose results are presented below.

In high-temperature superconductors, the hole pairing occurs in the vicinity of the copper-oxygen tetrahedra. In the SL based on lead chalcogenides, the aggregate of the available experimental data indicates that the electrons are paired at the dislocation sites. Hence the highest values of $T_c$ are possessed by $PbTe-PbS$ structures in which the separation \emph{t} between the MD network sites is the smallest ($t=52\ {\AA}$). For dislocation SL in $PbSe-PbS$ and $PbTe-PbSe$ at $T>1.5\ K$, only the initial segments of the SC transitions $R(T)$ are observed \cite{2}. The analogy observed in the structural elements and physical properties of HTS and SL of $PbTe-PbS$ (see below) leads to the assumption that according to the SC mechanism, the dislocation SL belongs to the class of HTS in spite of the low values of $T_c$ because of large values of \emph{t} and the different types of majority carriers (holes in HTS and electrons in SL). However, the latter circumstance is apparently not so important in view of the recent discovery of a new type of HTS with electron-type conductivity \cite{13}.

\section{PREPARATION AND STRUCTURE OF FILMS}

Samples were prepared in a vacuum unit with an oil-free pumping system ($P\sim10^{-4}-10^{-5}$\ Pa) by thermal vaporization of lead chalcogenides from tungsten "boats" followed by their subsequent condensation on the (001) surface of $KCl$ at a temperature 520-570\ $K$. The thickness of layers and the rate of condensation were controlled by a quartz cavity. Monolayer and two-layer films of $PbS$ and $PbTe$. The thickness of layers and the rate of condensation were controlled by a quartz cavity. Monolayer and two-layer films of $PbS$ and $PbTe$ were prepared as well as SL with the number of periods $N=1.5$ (three-layered sandwich $PbS-PbTe-PbS$) and $N=10$ and layer thickness $170-180\ \AA$. Preliminary investigations have shown that the SC properties of SL are determined not only by the thickness of the layers, but also by their stoichiometry to a considerable extent. Hence two series of SL with identical substructure but from different batches of $PbS$ were prepared, the first from a normal batch which gives a Hall carrier concentration $n_{H}\sim 10^{19} cm^{-3}$ and a mobility ${{\mu }_{H}}\sim {{10}^{2}}\ c{{m}^{2}}/V\cdot s$ (at 78~$K$) for monolayer films of thickness $5000\ \AA$ (denoted in the following by 78 $K$), and the second from a batch reduced to optimal composition by multiple sublimation \cite{14} which ensures the best stoichiometry for a given growth temperature, and having the parameters $n_{H}\sim 10^{17} cm^{-3}$, ${{\mu }_{H}}\sim {{10}^{4}}\ c{{m}^{2}}/V\cdot s$ (78~$K$) for monolayer samples of the same thickness (denoted by SL3, SL4).

In both series of SL, the $PbTe$ batch was reduced to optimal composition.

For structural investigations, the films were separated from the substrates by dissolving $KCl$ in distilled water and recovered on slides or the object screen of an electron microscope. The x-ray reflexes were photographed by the standard $\theta-2\theta$ scanning technique on the diffractometer DRON-2 in $Cu$  $K_{\alpha}$ radiation using a graphite (200) crystal as a secondary monochromator. Electron microscopic studies of the film structure were carried out on the electron microscope EM-100 AK with a resolution of $10\ \AA$.

Investigations revealed that lead chalcogenide films grow on one another according to the Frank-van der Merwe mechanism, resulting in the formation of monocrystalline layers with mosaic blocks repeating the blocks of the $KCl$ substrate (10-100 $\mu${m}). A regular square network of pure edge MD lying in the $\langle110\rangle$ direction with a Burgers vector \emph{a}/2 $\langle110\rangle$ and a period 52~\AA\ is formed on the (001) phase boundaries of $PbTe-PbS$ (see Fig.\ref{Fig1}a).

X-ray diffraction studies revealed that the SL are periodic with sharp interfaces as indicated by sharp reflex satellites around Bragg reflexes and near the primary beam. The SL period and the layer thickness were determined to a high degree of accuracy (to within 1 \AA) from the separation between satellite-reflexes \cite{15}.

\section{GALVANOMAGNETIC AND SUPERCONDUCTING PROPERTIES OF SUPERLATTICES}
Vital information about the galvanomagnetic and superconducting properties of SL was obtained from investigations of fluctuation conductivity, critical current, and upper critical fields, which were carried out by the four-probe method in a direct current in the temperature range $T=1.5-300~K$, and in magnetic fields up to 16~kOe \cite{2,3,4,5}. The value of $H_{c2}$ and the critical temperature $T_c$ were determined from the $0.5R_N$ level ($R_N$ is the sample resistance in the normal state). The volume-averaged Hall concentration of carriers in such SL was found to be $n_{H}\sim10^{19}~cm^{-3}$, and their mobility did not exceed $\mu_{H}=(300-600)~cm^{2}/V\cdot s$. It should be noted that the maximum carrier concentration is apparently observed in the vicinity of the heteroboundaries (this is associated with the presence of MD networks, see Fig. \ref{Fig1}a), while the carrier mobilities the lowest near these boundaries (the system is close to Anderson localization due to lattice distortions). The critical temperatures for the investigated SL were $3.5-3.9~K$ for the first series of samples, and $5.3-5.5~K$ for the second series.

According to the sign of the Hall constant, electrons are the main carriers in SL. Neglecting the contribution of holes in $PbS$ layers, the electron mean free path $l_i$ obtained by using the formula $l_i=(\hbar/e)\mu_{H}(3/4\pi^{2}n_{H})^{1/3}$ (Ref. \cite{16}) ($\mu_{H}$ is the carrier mobility and $e$ the electron charge) for lead chalcogenides is found to have the values presented in Table \ref{tab1}.
\begin{table}[]
\caption[]{}
\centering
\begin{tabular}{|c|c|c|c|c|c|} \hline
\ & \ & Carrier & \ &\ &\  \\
SL & No, of & conc., & $l_{i}$ ,\;{\AA}&  $\rho \cdot 10^{3} $ ,& $T_{c} ,\, K$ \\
number & periods & $n\cdot 10^{-19}$, & \ & $\Omega \cdot cm$ & \ \\
\  & \ & $cm^{-3}$ & \  & \ & \ \\ \hline
1 & 1.5 & 2.6 & 93 & 1.563 & 3.5 \\ \hline
2 & 10 & 1.3 & 285 & 0.806 & 3.5 \\ \hline
3 & 1.5 & 1.17 & 185 & 1.333 & 5.3 \\ \hline
4 & 10 & 1.55 & 368 & 0.556 & 5.5 \\ \hline
\end{tabular}
\label{tab1}
\end{table}

The magnitude of the quantity $l_i$ is found to be of the same order as the mean free path in $PbTe$ films of thickness 2000~\AA \ at low temperatures \cite{17}. In spite of the approximate nature of these estimates, it can perhaps be stated that at low temperatures $l_i$ becomes much larger than the coherence length $\xi_{\perp}$ at right angles to the layers, and is comparable in order of magnitude with longitudinal coherence length $\xi_{\parallel}$. This rules out the possibility of referring the SL data to the classical limit of dirty SC, although SL can be treated as quite "dirty" metals in view of the rather small value of the ratio ${l_i}/\lambda$, where $\lambda$ is the de Broglie wavelength of electrons (in our case, $\lambda\approx~100~\AA$).

The temperature and angular dependences of $H_{c2}$ (as well as the critical currents) of SL display singularities associated with a gradual localization (with decreasing temperature) of the order parameter, first in $PbTe$ Layers (3D-2\emph{D} crossover), and then on the MD network (2\emph{D}-2\emph{D'} transition) \cite{4,5}. The large anisotropy of magnetoresistance at low temperatures in the samples of the first series ${H_{c2}^{\parallel }}/{H_{c2}^{\bot }}\;\sim 5-7\ ({T}/{{{T}_{c}}\sim }\;0.6)$ as compared to the samples of the second series showing ${H_{c2}^{\parallel }}/{H_{c2}^{\bot }}\;\sim 2.5-3$ for the same $(T/T_{c})$, and a clearer observation of the 3\emph{D}-2\emph{D}-2\emph{D'} crossovers in the temperature and angular dependences of $H_{c2}$ lead to the conclusion that the SL are quasi-two-dimensional (weak coupling between layers) in the first series samples and anisotropically three-dimensional (strong coupling) in the second series samples.

\begin{figure}[]
\includegraphics[width=8cm,angle=0]{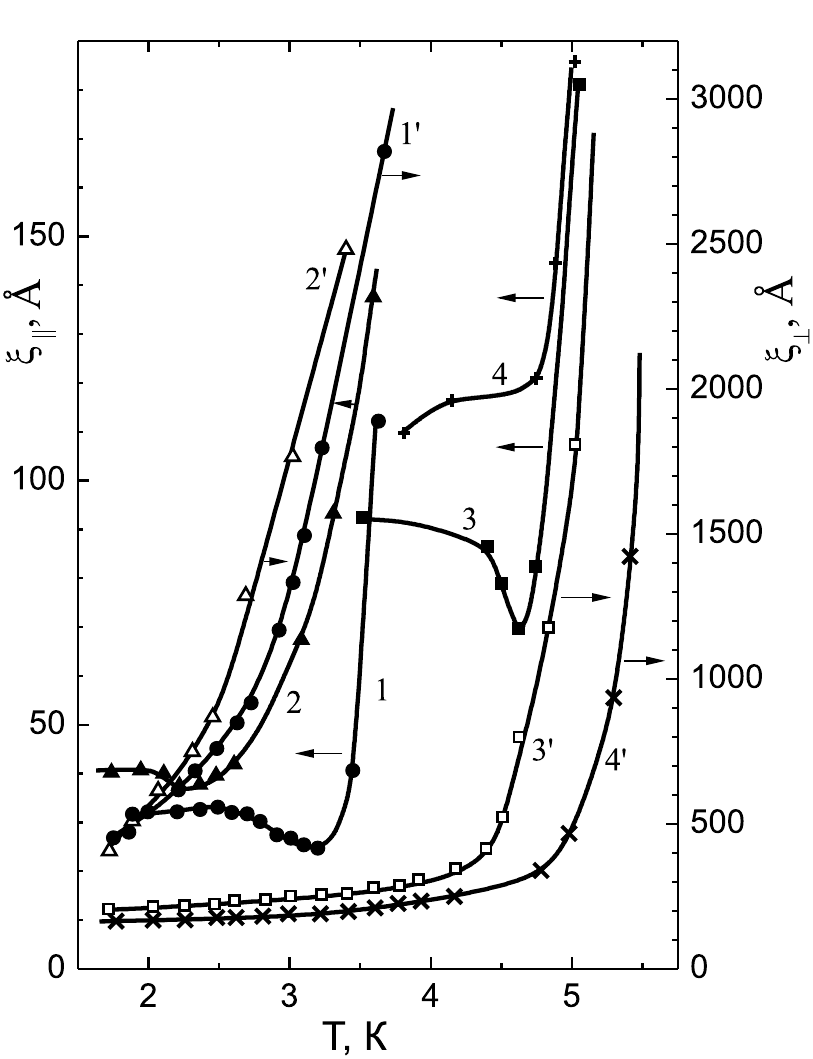}
\caption[]{Temperature dependence of the longitudinal $\xi_{\parallel}(T)$ and transverse $\xi_{\perp}(T)$  coherence lengths of superlattices SL1 (1, 1'), SL2 (2, 2'), SL3 (3, 3'), and SL4 (4, 4').}
\label{Fig2}
\end{figure}

Figure \ref{Fig2}
shows the temperature dependence of the longitudinal ${{\xi }_{\parallel }}(T)={{\left[ {{\Phi }_{0}}/2\pi H_{c2}^{\bot }(T) \right]}^{1/2}}$ (along the layers) and transverse ${{\xi }_{\bot }}(T)={{\Phi }_{0}}/2\pi {{\xi }_{\parallel }}(T)H_{c2}^{\parallel }(T)$ Ginzburg-Landau coherence lengths for samples of both series.

The nonmonotonic dependence and smallness of the quantities $\xi_{\perp}(T)$ at low temperatures is worth noting. For samples of both series, the minima of $\xi_{\perp}(T)$ correspond to temperatures at which the order parameter is localized in $PbTe$ layers \cite{4}. A displacement of the minimum $\xi_{\perp}$SL2(\emph{N}=10) towards lower temperatures compared to SL1 (\emph{N}=1.5) can be explained easily if we consider that SC is suppressed in a three-layer sandwich near the sample boundaries, corresponding to a localization of SC in the $PbTe$ layer starting from higher temperatures.

In the investigation of the fluctuational conductivity $\sigma$, two types of corrections ${\sigma}'={\sigma}(T)-{\sigma}_{0}$ are considered ${\sigma}(T)$ is the conductivity being measured, and ${\sigma}_{0}$ is the normal state conductivity), viz., the Aslamazov-Larkin (AL) correction \cite{18} taking into account the fluctuational formation of Cooper pairs, and the Maki-Thompson (MT) correction \cite{19,20,21} corresponding to the scattering of quasiparticles by superconducting pairs. Investigations of the fluctuational conductivity and magnetoresistance of layered systems including perovskite-type HTS provide an idea about the dimensions of these effects in different temperature regions above $T_c$, and hence about the degree of anisotropy of the samples under the investigation. Here, we proceed from the assumption that under conditions when the layered systems considered by us can be treated as quasi-two-dimensional, the estimates of fluctuational quantities used for two-dimensional systems are qualitatively acceptable \cite{18}.

The dependence of the normalized excess conductivity ${\sigma}'/{\sigma}_0$ for $T>T_c$ on the reduced temperature ${\tau}=(T-T_{c})/T_c$, plotted in logarithmic coordinates makes it possible to use the AL theory to determine the dimensions of fluctuations from the slope of the linear segments, since in this theory we have ${\sigma}'/{\sigma}_{0}\sim {\tau}^{-k}$, where the exponent \emph{k}=0.5, 1, and 2 for three-, two-, or zero-dimensional fluctuations respectively.

\begin{figure}[]
\includegraphics[width=8cm,angle=0]{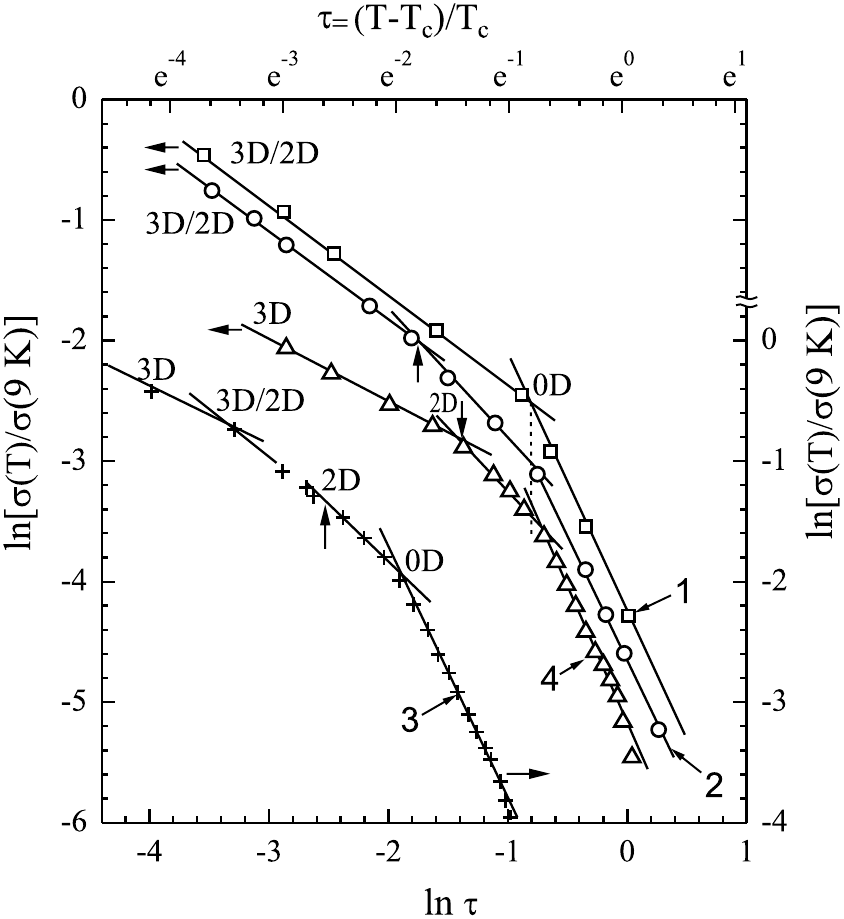}
\caption[]{Dependence of the normalized excess conductivity on the reduced temperature $\tau$ in logarithmic coordinates: 1) SL1 (see inset to Fig.\ref{Fig5}), 2) the same in a magnetic field $H_{\parallel}=500~Oe$, 3) dependence constructed for $Cu$-SL3 point contact (Figs. \ref{Fig9},\ref{Fig10} by using the temperature dependence $R_{D}^{0}(T)$,  4) the dependence constructed for \emph{Cu}-SLl point contact (Figs.\ref{Fig5},\ref{Fig8}) by using the temperature dependence $R_{D}^{V}(T)$.}
\label{Fig3}
\end{figure}

Investigations of the fluctuational region of SL for two series with $T_{c}\sim 3.5$ and $5.3~K$ show (Fig. \ref{Fig3} that the excess conductivity in a magnetic field or without it can be described qualitatively by the AL quantum correction. As in the case of HTS \cite{22},  the insignificant contribution of the MT correction apparently points towards the small values of $\tau_{\varphi}$, the phase disturbance time for the wave function of quasiparticles.

The main peculiarities of the temperature dependence of fluctuational conductivity can be studied by considering the example of a three-layer sandwich SL1 (Fig.\ref{Fig3}). It can be seen that, as the temperature is lowered in the absence of a magnetic field, a linear region is observed from $\tau \sim 1$ with slope \emph{k}=2 corresponding to the zero-dimensional fluctuational conductivity. This is followed by a ($3D/2D$) region with slope \emph{k}=3/4 intermediate between \emph{k}=1/2 and \emph{k}=1, which can be conditionally called quasi-two-dimensional. Application of a magnetic field $H_{\parallel}=500~Oe$ practically does not alter the boundary of the zero-dimensional region, but is followed by a region of purely two-dimensional fluctuations (\emph{k}=1) which is replaced by a region of quasi-two-dimensional fluctuations. An analysis of the crossover temperature from one type of ${\sigma}(T)$ dependence to another shows that zero- dimensional fluctuations appear for $\tau=1$ and terminate at $\tau\simeq{(2t/t_{PbTe})}^2$.  In a field $H_\parallel$ the region of quasi-two-dimensional fluctuations begins from $\tau\simeq{(\sqrt{2}\ t/t_{PbTe})}^2$ (corresponding values of $\tau$ are marked by arrows in Fig.\ref{Fig3}). Such a behavior of $\sigma'(T)$ can be explained by assuming that there exists a certain coherence length $\xi(T)=\xi_{0}/\sqrt{\tau}$ whose approximate equality to the size of the characteristic structural elements changes the dimensionality of fluctuations. Assuming that $\xi_{0}=t$ viz., the separation between MD network sites, the beginning of the zero-dimensional fluctuations corresponds to the temperature at which $\xi(T)=t,\ \tau=1 \ (T=2T_c)$. In this case, a transition to two-dimensional fluctuations occurs $(H=0)$ at $\xi(T)\simeq(t_{PbTe}/2)$, when the stabilization of the two-dimensional superconductivity begins as a result of the interaction through the $PbTe$ layer. The application of a magnetic field effectively decreases the dimensionality of fluctuations, hence the zero-dimensional region is followed by a region of purely two-dimensional fluctuations which is transformed into a quasi-two-dimensional region for $\xi(T)\simeq(t_{PbTe}/\sqrt {2})$ (this is analogous to the $2D-3D$ crossover in layered compounds, where the role of the "insulating" inter-layer is played by $PbTe$). The absence of a purely three- dimensional region can be attributed to the fact that the order parameter for a sandwich is concentrated mainly in the $PbTe$ layer and attains its maximum value in the immediate vicinity ($\sim t$) of the MD networks. In other words, the SC state in the investigated temperature range is highly nonuniform over the sample thickness. The slight departure of the experimental value of  $\ln\tau$ for the $0D-(3D/2D)$ crossover point from $\ln[{(2t/t_{PbTe})}^2]$ to higher module values is probably due to the closely situated sample boundaries, which "push" the order parameter towards the middle of the sample. For want of space, we shall not discuss in detail the peculiarities of the "bulk" samples of SL with $N$=10, but simply observe that for "bulk" SL also, the points corresponding to the change in the dimensionality of fluctuations obey equally simple relations although the transitions between segments are not so sharp and are more blurred in temperature.

Although the coherence length $\xi_0$ introduced formally by us makes it possible to describe quite well the positions of the points corresponding to a change in the dimensionality of fluctuations, its physical meaning must be explained. An analysis of the set of superlattices with different thicknesses of $PbTe$ and $PbS$ layers, different mobilities and electron concentrations showed that the maximum value of $T_c$ is attained in structures in which the unit cell of the three-dimensional SL (of size $\Lambda \times t \times t$) contains about 18 electrons. This circumstance can be explained by the fact that all these electrons are mainly localized in the vicinity of dislocation sites, and each lead atom at a site marked by a dark circle in Fig. \ref{Fig1}a (see the middle figure) has two electrons while the lead atom at the base of the tetrahedron formed by S atoms (light circle) has one "unpaired" electron. Considering that each unit cell of SL has two dislocation sites, we obtain 18 electrons per cell. Assuming that pairing occurs in the vicinity of a dislocation site as a result of interaction of "unpaired" electrons at the adjacent sites of the same dislocation network, the natural minimum coherence length for which this process is possible is the separation \emph{t} between sites. Eighteen electrons per unit cell correspond to the average number of pairs per lattice site, which is equal to 1/2 (this is the optimal value for attaining the highest value of $T_c$ in Kulik's theory of local pairs \cite{23} in the mean field approximation). It should be noted that for systems $YBaCuO$ and $Bi(Sr, Ca)CuO$, the peak value of $T_c$ is also observed for the same number of mobile holes per copper atom in the $CuO_2$ plane (about 0.45 per unit cell of $YBaCuO$) \cite{24}. The nonintegral value of this quantity may be due to correlation effects \cite{25}.

In this model, the onset of fluctuational pairing must indeed correspond to the correlation length $\xi(T)=t$. When pairs in adjacent networks begin to interact through the $PbTe$ layer ($\xi=t_{PbTe}/2)$, the SC fluctuations can propagate along the networks and are of quasi-two-dimensional nature. The application of a magnetic field probably leads to the "expulsion" of the order parameter from inner regions of $PbTe$ to dislocation networks. This results in the emergence of a 2\emph{D}-region which is replaced upon a further increase in $\xi(T)$ by a quasi-two-dimensional region for $\xi>t_{PbTe}/\sqrt{2}$ (in the theory of layered SC, the $3D-2D$ dimensional crossover occurs for $H_{c2}$ under the same condition, except that $t_{PbTe}$ is replaced by the separation between SC planes). The absence of the SC transition in the two-layer sandwich $PbTe-PbS$ mentioned above speaks in favor of the hypothesis of pairing on a dislocation network since it is well known that phase fluctuations of the order parameter in a two-dimensional system violate the long-range order which can be restored even under a weak interaction with at least one analogous system.

It should be also observed that we compare $\xi$ with characteristic layer thicknesses. In other words, $\xi$ has the meaning of the coherence length $\xi_\bot$. The Ginzburg-Landau coherence lengths $\xi_\bot(0)$ for $YBaCuO$, calculated from fluctuational conductivity and from temperature dependence of $H_{c2}$ differ by a factor of two. The value of $\xi_0$ introduced by us is also clearly different from the low-temperature values of $\xi_\bot$ determined from the data for $H_{c2}$ (see Fig.\ref{Fig2}). The reason behind such a discrepancy is not known at present.

A dependence similar to that for the case $H_\parallel\neq0$ was also observed for the differential resistance of an SL-\emph{Cu} point contact for high bias voltages across the contact ($eV\gg\Delta_0)$  $R_{D}^{V}(T)$ (see below for details). However, a transition to the two-dimensional $(3D-2D)$ region begins at a temperature slightly higher than $\tau={(\sqrt{2}\ t/t_{PbTe})}^2$. This may be associated with the peculiarities of current flow in the vicinity of the contact and with the effect of the magnetic field of the current on the SC fluctuations. Instead of the region $3D/2D$ (for the case $H_\parallel\neq0$), we now obtain a three-dimensional region.

\begin{table*}[]

\caption[]

\begin{tabular}{|c|c|c|c|c|c|c|c|}\hline
Fluctua- & \ &  \multicolumn{2}{|c|}{SL1}& SL1 & SL3 & $YBaCuO$ \cite{39}* & $YBaCuO$  \\ \cline{3-7}
tion & Para- & \multicolumn{2}{|c|} {Resistive meas} & \multicolumn{3}{|c|} {PC meas. using:} & (Ref. \cite{22,27}) \\ \cline{3-7}
dimen. & metres & \multirow{2}{*}{$H=0$} & \multirow{2}{*}{$H_{\parallel}=500~Oe$} & \multirow{2}{*}{$R_{D}^{V}(T)$} & \multirow{2}{*}{$R_{D}^{0}(T)$} & \multirow{2}{*}{$R_{D}^{0}(T)$}& resistive \\
\ & \ & \ &\ & \ & \ & \ &measurements \\ \hline \hline

\multirow{5}{*}{$3D$}& $T,~K$ & - & - & {\footnotesize $3.8<T<4.4$} & {\footnotesize $5.4<T<5.5$} & {\footnotesize $101<T<120$} & {\footnotesize $94<T<100$} \\ \cline{2-8}
& $\tau$ & - & - &{\footnotesize $0.086<\tau <0.257$} &{\footnotesize $0.02<\tau <0.038$} &{\footnotesize  $0.123<\tau <0.333$} &{\footnotesize  $0.044<\tau <0.111$} \\ \cline{2-8}
& ${T}',\ K$& - & - &{\footnotesize  $3.81<{T}'<4.16$} & {\footnotesize $5.42<{T}'<5.53$} & {\footnotesize $99.96<{T}'<119.9$} &{\footnotesize  $92.49<{T}'<99.96$} \\ \cline{2-8}
& ${\tau }'$ & - & - &{\footnotesize  $0.088<{\tau }'<0.187$} & {\footnotesize $0.022<{\tau }'<0.044$} & {\footnotesize $0.111<{\tau }'<0.332$} & {\footnotesize $0.028<{\tau }'<0.111$}  \\ \cline{2-8}
 & $\xi $ & - &- & {\footnotesize $\Lambda /2>\xi >{t}'/\sqrt{2}$} & {\footnotesize $\Lambda >\xi >\Lambda /\sqrt{2}$} & {\footnotesize $\Lambda /2>\xi >{t}'$}  & {\footnotesize $\Lambda >\xi >\Lambda /2$} \\ \hline \hline

\multirow{5}{*}{$\frac{3D}{2D}$}& $T,~K$ & {\footnotesize $T<5$} & {\footnotesize $T<4.11$} & - & {\footnotesize $5.5<T<5.8$} & - & - \\ \cline{2-8}
& $\tau$ & {\footnotesize $\tau <0.429$} & {\footnotesize $\tau <0.1752$} & - & {\footnotesize $0.035<\tau <0.094$} & - & - \\ \cline{2-8}
& ${T}',\ K$& {\footnotesize ${T}'<4.81$} & {\footnotesize ${T}'<4.16$} & - & {\footnotesize $5.53<{T}'<5.77$} & - & - \\ \cline{2-8}
& ${\tau }'$ & {\footnotesize ${\tau }'<0.374$} & {\footnotesize ${\tau }'<0.1871$} & - & {\footnotesize $0.044<{\tau }'<0.088$} & - & - \\ \cline{2-8}
& $\xi $ & $\xi >{t}'/2$ & {\footnotesize $\xi >{t}'/\sqrt{2}$} & - & {\footnotesize $\Lambda /\sqrt{2}>\xi >\Lambda /2$} & - & - \\  \hline \hline

\multirow{5}{*}{$2D$}& $T,~K$ & - & {\footnotesize $4.11<T<5$} &  {\footnotesize $4.4<T<5$} & {\footnotesize $5.8<T<6.1$} &  {\footnotesize $120<T<180$} & {\footnotesize $100<T<125$} \\  \cline{2-8}
& $\tau$ & - & {\footnotesize $0.175<\tau <0.429$} & {\footnotesize $0.257<\tau <0.429$} & {\footnotesize $0.094<\tau <0.15$} & {\footnotesize $0.333<\tau <1$} & {\footnotesize $0.111<\tau <0.389$} \\  \cline{2-8}
& ${T}',\ K$& - & {\footnotesize $4.16<{T}'<4.81$} & {\footnotesize $4.16<{T}'<4.81$} & {\footnotesize $5.77<{T}'<6.29$} & {\footnotesize $119.9<{T}'<180$} & {\footnotesize $99.96<{T}'<119.87$} \\  \cline{2-8}
& ${\tau }'$ & - & {\footnotesize $0.187<{\tau }'<0.374$} & {\footnotesize $0.187<{\tau }'<0.374$} & {\footnotesize $0.088<{\tau }'<0.19$} & {\footnotesize $0.332<{\tau }'<1$} & {\footnotesize $0.111<{\tau }'<0.332$} \\  \cline{2-8}
& $\xi $ & - & {\footnotesize ${t}'/\sqrt{2}>\xi >{t}'/2$} & {\footnotesize ${t}'/\sqrt{2}>\xi >{t}'/2$} & {\footnotesize $\Lambda /2>\xi >{t}'/\sqrt{2}$} & {\footnotesize ${t}'>\xi >t/2$} & {\footnotesize $\Lambda /2>\xi >{t}'$}  \\ \hline \hline

\multirow{5}{*}{$0D$}& $T,~K$ & $5<T<7$ & $5<T<7$ & $5<T<7$ & $6.1<T<7.1$ & $180<T<220$ & $125<T<180$ \\  \cline{2-8}
& $\tau$ & $0.429<\tau <1$ & $0.429<\tau <1$ & $0.429<\tau <1$& $0.151<\tau <0.34$ & $1<\tau <1.44$ & $0.389<\tau <1$ \\  \cline{2-8}
& ${T}',\ K$& $4.81<{T}'<7$ & $4.81<{T}'<7$ & $4.81<{T}'<7$ & $6.29<{T}'<7.28$ & $180<{T}'<210$ & $119.87<{T}'<180$\\  \cline{2-8}
& ${\tau }'$ & $0.374<{\tau }'<1$ & $0.374<{\tau }'<1$ & $0.374<{\tau }'<1$ & $0.187<{\tau }'<0.37$ & $1<{\tau }'<1.328$ & $0.332<{\tau }'<1$  \\  \cline{2-8}
& $\xi $ & ${t}'/2>\xi >t$ & ${t}'/2>\xi >t$ & ${t}'/2>\xi >t$ & ${t}'/\sqrt{2}>\xi >{t}'/2$ & $t/2>\xi >{t}'/2$ & ${t}'>\xi >t/2$ \\ \hline
\end{tabular}

\footnotesize
\vspace{5pt}
\begin{flushleft}\textbf{Remark.} Here, ${T}'$  and ${\tau }'$  are calculated values: $\xi (T')={{\xi }_{0}}/\sqrt{{{\tau }'}};$ $\xi =L\to {\tau }'={{\left( {{{\xi }_{0}}}/{L}\; \right)}^{2}}$ ($L$ is the characteristic size of the system).\\The following values of the parameters were used in computations: for SL (see Fig. \ref{Fig1}a)\ \  ${t}'={{t}_{PbTe}}=170~\AA$;\ \  $\Lambda =350~\AA;$ ${{\xi }_{0}}=t=52~\AA$ $T_{c} (SL1) = 3.5~K$,
  $T_c$(SL3)=5.3~$K$; for $YBa_{2}Cu_{3}O_{6.91}$ for Ref.\cite{51} (see Fig.\ref{Fig1}b) $t'=3.3676~\AA $,  $\Lambda =C=11.666~\AA$, $t=b=3.88~\AA$, ${{\xi }_{0}}=t/2=1.94~\AA$, $T_{c}=90~K$.
 The dependence of the unit cell size on oxygen concentration must be taken into consideration while analyzing the HTS samples.\\

\textbf{*}The upper boundary of the temperature interval for zero-$D$ fluctuations was determined from the temperature corresponding to the onset of metal-insulator transition by analyzing the shape of $dV/dI$ for point contacts.
\end{flushleft}
\normalsize
\label{tab2}
\end{table*}

The behavior of samples with high $T_c$ differs sharply from that of quasi-two-dimensional samples. The beginning of the zero-dimensional region (see (+) in the plot shown in Fig.\ref{Fig3}, constructed by using the temperature dependence $R_{D}^{0}(T)$ of the differential resistance of the \emph{Cu}-SL3 point contact for $V=0$) is displaced towards lower $\tau={(2t/t_{PbTe})}^2$. In other words, the zero-dimensional fluctuations are manifested for $\xi<(t_{PbTe}/2)$. Subsequent transitions ($0D-2D-3D/2D-3D$) occur for $\xi(T)\simeq(t_{PbTe}/{\sqrt{2}})$, $(\Lambda/2)$, $(\Lambda/\sqrt{2})$. The displacement of the onset of the zero-dimensional fluctuation region towards lower $\tau$ indicates that the initial pairing of electrons for SL with large $T_c$ (and $l_i$) may occur not in the plane of dislocation networks (as in quasi-two-dimensional samples), but between networks. Such a possibility (leading to an increase in $T_c$) was predicted, for example, in Ref.\cite{26} for layered compounds and was attributed to the peculiarities of screening of Coulomb interaction of electrons in mixed conducting layers.

An analysis of the data available in literature for the $YBaCuO$ family (see, for example, Refs.\cite{22,27,28}) shows{\renewcommand{\thefootnote}{*}\footnote{The data for $Bi$ and $Tl$ families of HTS are too unreliable and sketchy at present for drawing any conclusions on the behavior of fluctuational conductivity.}}
 that we can isolate for these materials regions of 0\emph{D}-, 2\emph{D}-, and 3\emph{D}-fluctuational conductivity, the zero-dimensional fluctuations{\renewcommand{\thefootnote}{**}\footnote{As a rule, publications on HTS contain data for two- or three-dimensional fluctuation regions. To our knowledge, the only publication in which the zero-dimensional region is clearly followed in a $YBa_{2}Cu_{3}O_x$ film with $T_c=90~K$ is Ref. \cite{27}.}\addtocounter{footnote}{-1}} beginning from $\tau\simeq 1$, two-dimensional from $\tau\simeq{(t/2t')}^2$, and three-dimensional from $\tau\simeq{(t/\Lambda)}^2$, where $t'$ is the separation between $CuO_2$ planes (see Fig.\ref{Fig1}b) and $\Lambda$ is the lattice parameter along the \emph{c}-axis. In the present case, $\xi_0$ equal to $t/2$, i.e., the distance between $Cu$ and $O$ atoms in $CuO_2$ planes (\emph{t} is the larger lattice parameter in the basis plane \emph{ab}, $t=b$). Then the two-dimensional region begins at $\xi(T)\simeq t'$, and the three-dimensional region at $\xi\simeq\Lambda/2$. Hence it can be naturally assumed that pairs are created in the vicinity of oxygen-copper pyramids in the $CuO_2$ plane, and for pairs of size approximately equal to $t'$, two-dimensional fluctuational conductivity sets in on twinned $CuO_2$ planes. Magnetic fields displace the $2D-3D$ transition towards lower temperatures $(\xi(T)=\Lambda)$. This can be seen clearly, for example, in Fig.\ref{Fig3} from Ref.\cite{28} for an $Er$ ceramic whose behavior differs from that of $YBaCuO$ (well-defined two-dimensional regions were observed only in magnetic fields; for $H=0$ the departure from the 3\emph{D}-dependence started at $\xi<\Lambda/2$. The value $\xi_0=t/2$ matches well with the Ginzburg-Landau correlation length $\xi_\perp(0)$ for $YBa_{2}-Cu_{3}O_{7-\delta}$ calculated from the fluctuational conductivity in many works (in Ref.\cite{22}, $\xi_\perp=2.0\pm0.5~\AA$). For the sake of convenience in orientation, the results obtained in the present work and the data of PC investigations of $YBa_{2}-Cu_{3}O_{7-\delta}$ (Ref.\cite{39}) are compiled in Table \ref{tab2} which also shows the values of $\tau$ for which a transition to the region of critical fluctuations takes place.

The above results show that the behavior of the fluctuational conductivity of SL $PbTe-PbS$ is quite similar to that of HTS materials. However, the geometrical parameters of SL1-SL4 differ strongly from the parameters of the $YBaCuO$ family, since the layer thickness in them considerably exceeds the separation between the dislocation network sites while, on the contrary, the separation between $Cu$ atoms in $YBaCuO$ is larger than the separation $d'$ between planes. Such a situation could be modeled on an SL with quite thin layers, but unfortunately the geometry of dislocation networks varies radically for $t_{PbTe}<t$, adjacent networks are displaced relative to one another by $t/2$, and a different situation emerges in which the sites of the network are not situated one below the other. The situation in an HTS and an SL with different layer thicknesses will be discussed in detail in a separate communication.

Apart from the traditional methods of investigation of SC properties, vital information on the nature of order parameter localization and on the SC energy gaps of SL can be obtained from PC investigations.

\section{EXPERIMENTAL TECHNIQUE}
The current-voltage characteristics (IVC) as well as their first and second derivatives for spring point-contacts $Cu$-SL were measured at right angles to the layers. The copper electrode was in the form of a trihedral pyramid cut from a single crystal on an electric erosion setup. Details of the conditions of electrochemical treatment and the quality control of the surface of the copper electrode are presented in Ref. \cite{29}. The radius of rounding of the prism tip was about $3-5~\mu m$.

In order to form the point contact, the tip of the pyramid was pressed with a certain force against the plane surface of the SL, and then moved along the surface. The temperature measurements were carried out in an intermediate cryostat with a capillary \cite{30}.

\section{DISCUSSION OF EXPERIMENTAL RESULTS}
\subsection{Location of bridges in the bulk of heterostructures}

By changing the compression of the copper electrode against the SL surface, we can form a bridge in the upper layer or in the bulk of a heterostructure. For the given geometry of the experiment, the use of a tunnel microscope in the point-contact spectrometer regime would not have any special advantages over the traditional point-contact spectroscopy. As a needle is pressed into a heterostructure, its displacement can be determined from the controlling voltage across the piezoelectric element. Apparently, for a given voltage this displacement differs from the displacement in vacuum and depends on the geometrical size of the tip of the needle, as well as the microhardness of the heterostructure. Even if these factors are somehow taken into account and the microhardness of the needle is much more than of the material under investigation so that it penetrates to a certain depth without any noticeable plastic deformation, the formation of a bridge at a given depth cannot be ruled out. The needle is always covered with an oxide layer, and the "weak" spot on the surface where the oxide layer is violated and a bridge is formed may exist at any depth right up to its maximum value equal to the thickness to which the tip of the needle penetrates the material.

However, since SL1 and SL2 have a quasi-two-dimensional superconductivity, we can use the traditional spectroscopic technique for point contacts to determine the possible spot of formation of a bridge in a heterojunction from the temperature dependence of the excess current. Superconductivity is induced in the entire volume of the SL only at temperatures close to $T_c$ and disappears first in the $PbS$ layer as the temperature is lowered, shrinking ultimately to the heteroboundaries.

\begin{figure}[]
\includegraphics[width=8cm,angle=0]{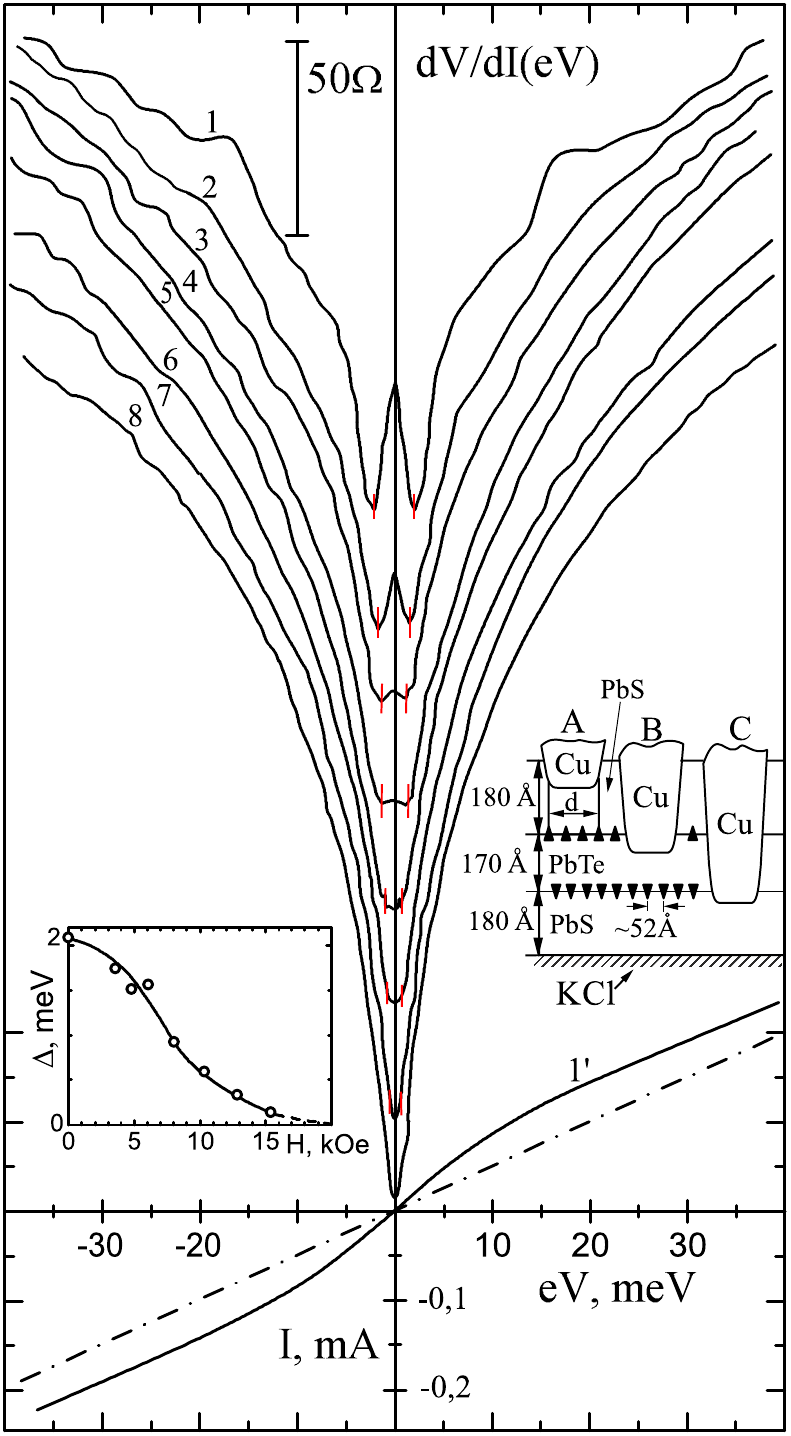}
\caption[]{IVC and their first derivatives for the point contact $Cu$-SL1 at $T=1.85~K$ under magnetic fields $H$=0(1, 1'), 3.8(2), 5.1(3), 6.25(4), 7.9(5), 10.5(6), 13(7) and 15.6(8) $kOe$. The dot-dash line is parallel to the IVC for $eV\gg\Delta$. The inset to the left shows the dependence of the energy gap on the magnetic field, while the right-hand inset shows the structure of SL1 and various models of the point contact.}
\label{Fig4}
\end{figure}

Figures \ref{Fig4} and \ref{Fig5} show the characteristics of point contacts formed perhaps in the immediate vicinity of the superconducting heteroboundary in the $PbTe$ layer (model $B$ in the inset to Fig.\ref{Fig4}). This is indicated by the existence of an excess current right up to the lowest temperatures. For the point contacts shown in Figs.\ref{Fig6} and \ref{Fig7}, there is no excess current at low temperatures. It can be seen from the first IVC derivative that the excess current for the contact in Fig.\ref{Fig6} appears only for $T>2.54~K$. Apparently, this contact is formed in the $PbS$ layer (model $A$ in the inset to Fig.\ref{Fig4}) at a distance larger than $\xi_{\perp}(0)$ from the heteroboundary.

\subsection{Temperature dependence of the differential resistance of a point-contact for $eV\gg\Delta$. Use of point-contacts for determining the dimensionality of superconducting fluctuations. Superconductivity and localization}

\begin{figure}[h!]
\includegraphics[width=8cm,angle=0]{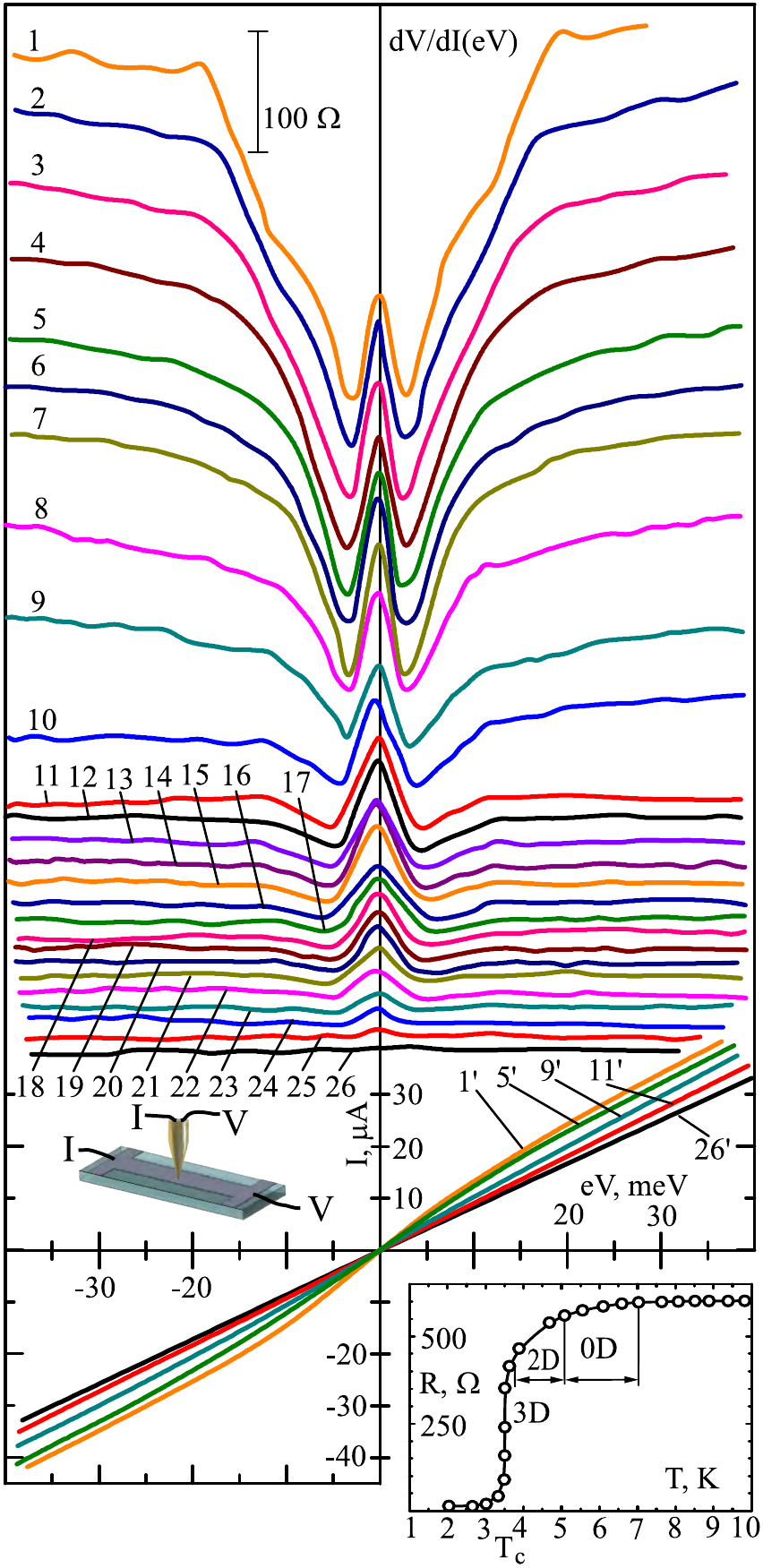}
\caption[]{IVC and their first derivatives for the point-contact $Cu$-SL1 at $H$=0 and $T$=1.85(1, 1'), 2.01(2), 2.2(3), 2.36(4), 2.6(5, 5'), 2.8(6), 3(7), 3.28(8), 3.6(9, 9'), 3.8(10), 4(11, 11'), 4.2(12), 4.4(13), 4.6(14), 4.8(15), 5.0(16), 5.2(17), 5.4(18), 5.6(19), 5.8(20), 6.0(21), 6.2(22), 6.4(23), 6.6(24), 6.8(25), and 7(26, 26') $K$. The inset to the left shows the experiment geometry, while that on the right-hand side shows the temperature dependence of the resistance of SL1.}
\label{Fig5}
\end{figure}

For the series of contacts studied by us, the temperature and magnetic field dependences of the differential resistance $R_{D}^{V}$ were measured for $eV\gg\Delta$. The $R_{D}^{V}(T)$ or $R_{D}^{V}(H)$ dependences were measured not only for contacts with a direct conductivity (Figs.\ref{Fig5}, \ref{Fig9}), but also for purely tunnel-type point contacts (Fig.\ref{Fig7}). Figure \ref{Fig8} shows the $R_{D}^{V}(T)$ dependence for bias voltages
\begin{figure}[]
\includegraphics[width=8cm,angle=0]{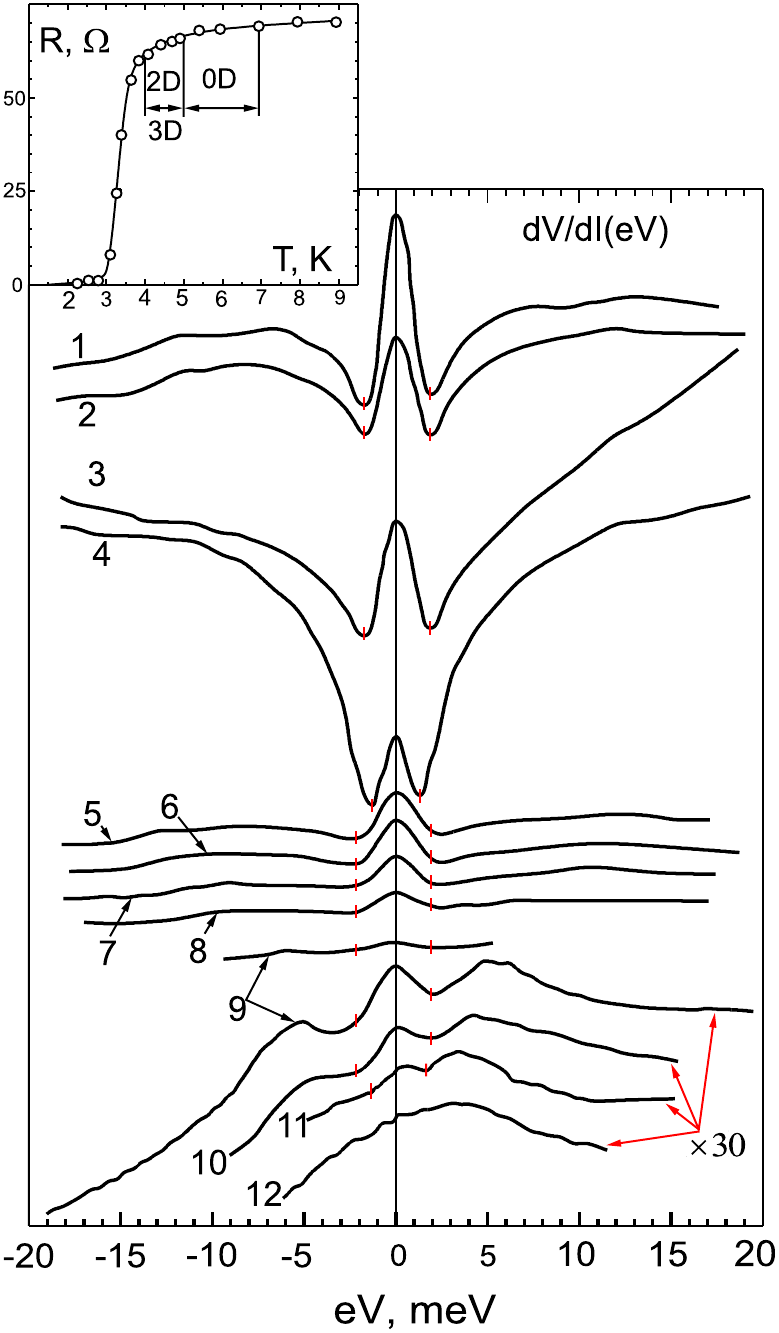}
\caption[]{IVC and their first derivatives for the point contact $Cu$-SL1 for $H=0$ $R_{D}^{0}(1.75\ K)=190\ \Omega $ and $T$=1.75(1), 1.98(2), 2.54(3), 3.2(4), 4.2(5), 4.9(6), 5.4(7), 6(8), 6.7(9), 6.8(10), 6.9(11) and 7(12) $K$. The inset shows the temperature dependence of the resistance of SL2.}
\label{Fig6}
\end{figure}
\begin{figure}[]
\includegraphics[width=8cm,angle=0]{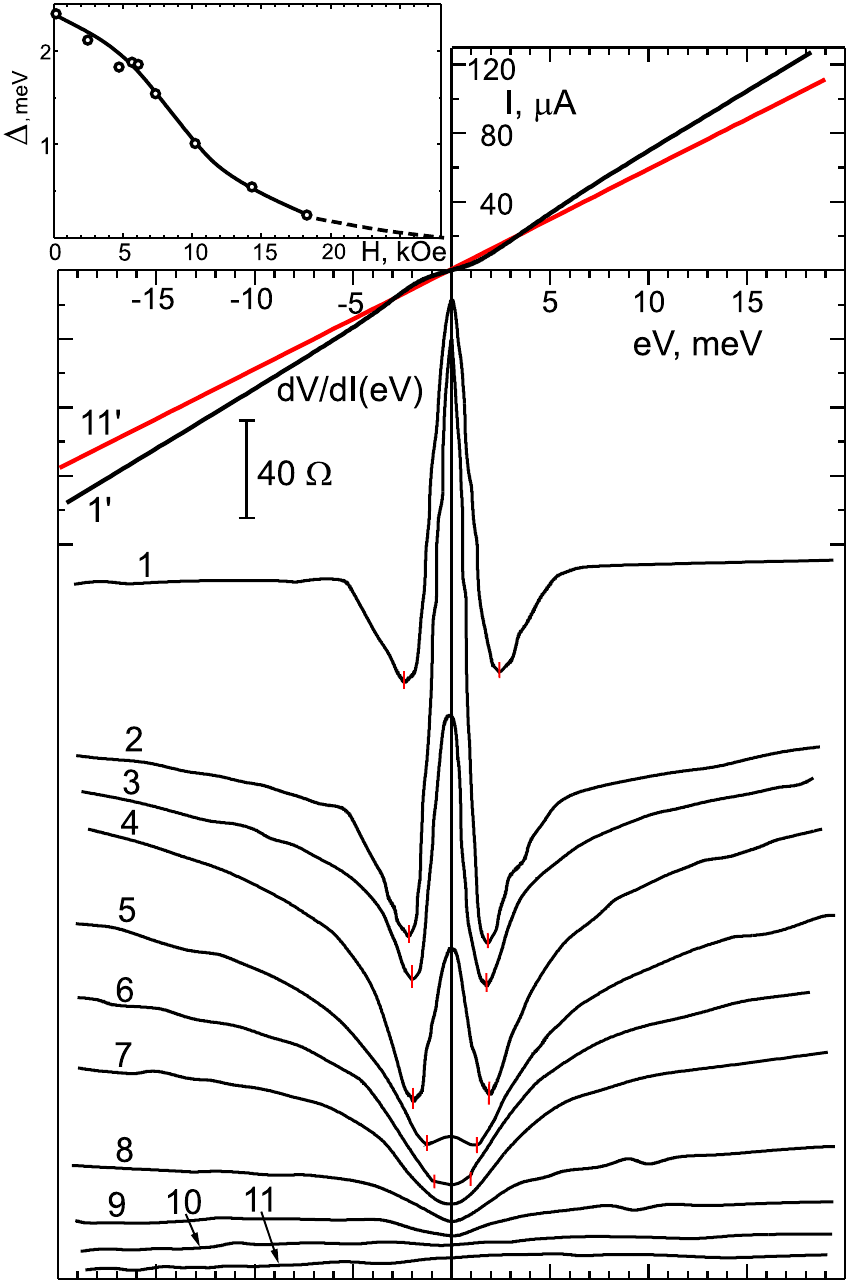}
\caption[]{IVC and their first derivatives for the point contact $Cu$-SL2 at $T$=1.84  $K$ and $H$=0(1, 1'), 2.5(2), 4.6(3), 5.8(4), 7.4(5), 9.4 (6), 10.4(7), 14.3(8), 18.2(9), 23.4(10), and 29.9(11, 11') $kOe$. The inset shows the dependence of the energy gap on the magnetic field.}
\label{Fig7}
\end{figure}

\begin{figure}[]
\includegraphics[width=8cm,angle=0]{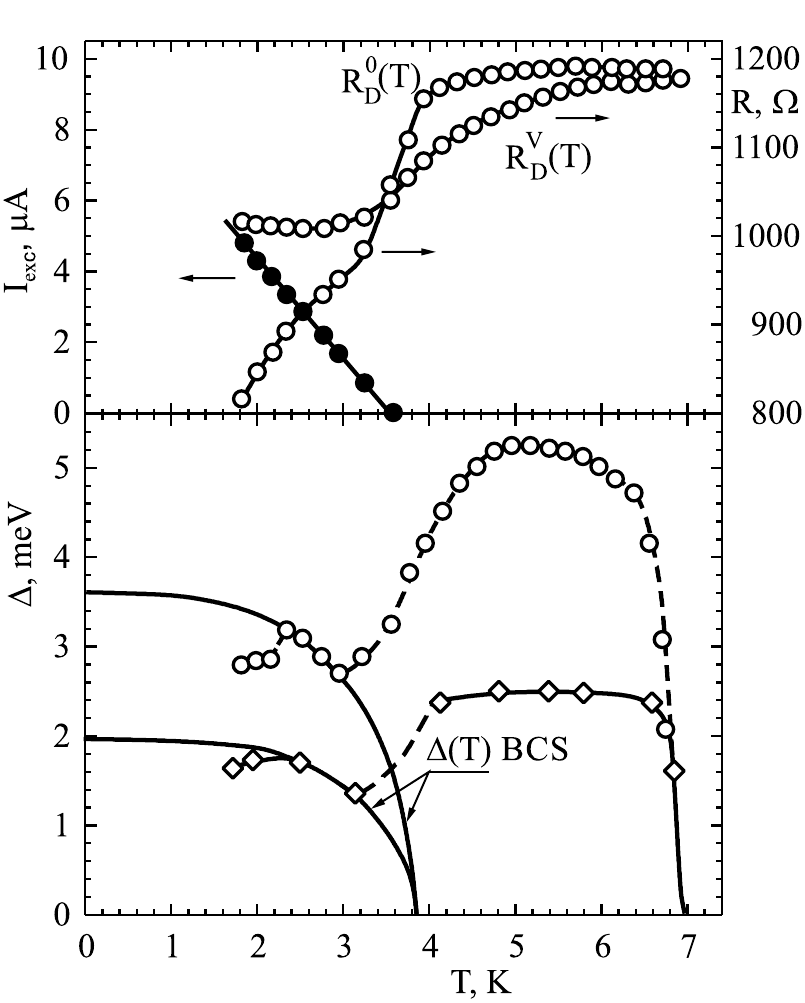}
\caption[]{Temperature dependence of the energy gaps (the dark circle and square correspond to the points of coincidence with the BCS dependences), excess current $I_{exc}$ and the differential resistance $R_{D}^{V}(T)$ for zero and $R_{D}^{0}(T)$ for large bias voltages across the contact ($\circ$ and $\bullet$ shows the contact in Fig. \ref{Fig5}, and $\diamond$ the contact in Fig.\ref{Fig6}). The value $T_{c}\sim 3.9~Ê$  corresponds to the onset of the SC transition.}
\label{Fig8}
\end{figure}

\begin{figure}[]
\includegraphics[width=8cm,angle=0]{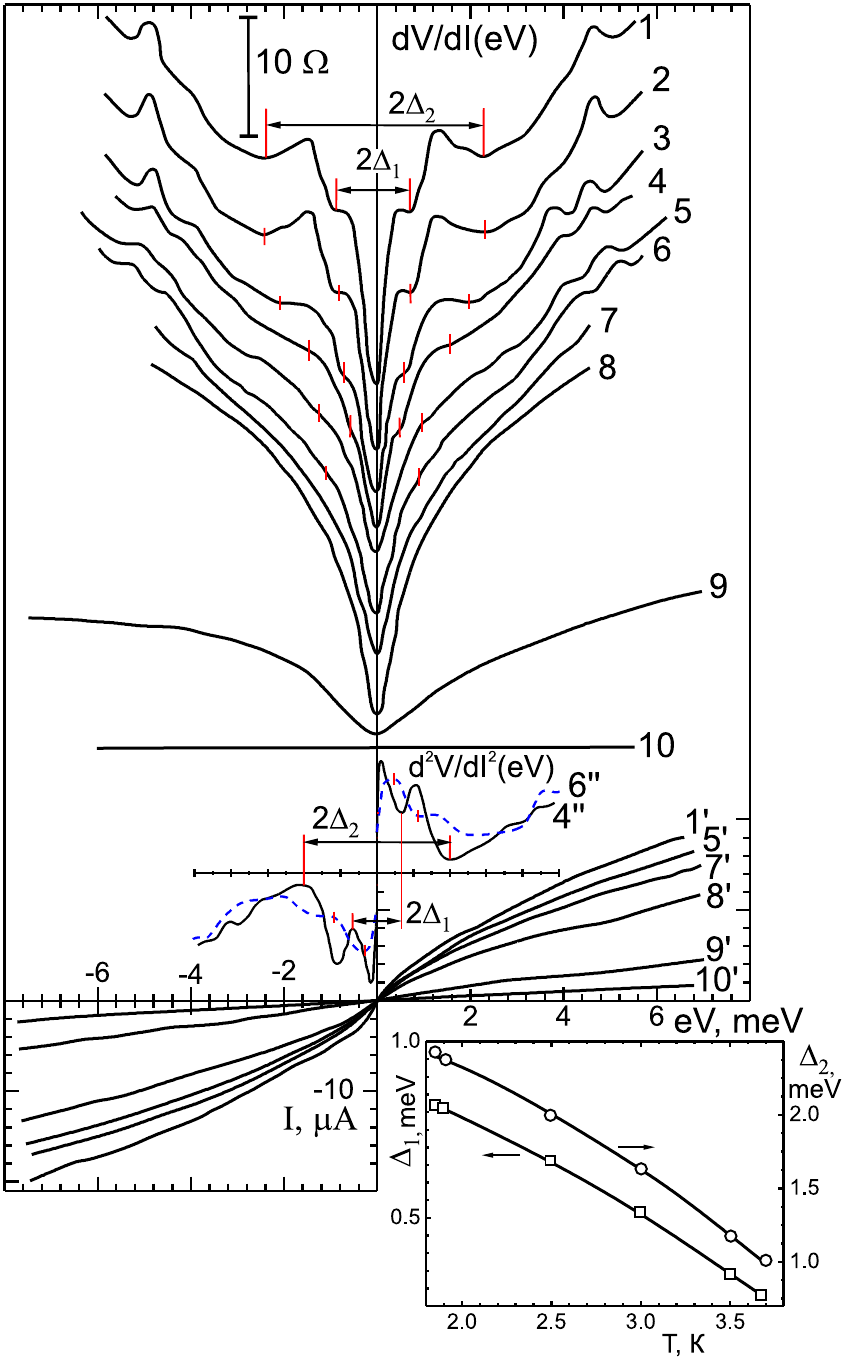}
\caption[]{IVC and their first and second derivatives for the point contact $Cu$-SL3 for $H$=0 and $T$=1.86(1, 1'), 2(2), 2.5(3), 3(4, 4''), 3.5(5, 5'), 3.7(6, 6''), 4(7, 7'), 4.5(8, 8'), 5 (9,9'), and 8(10, 10')~$K$. The methods of determining the gaps $\Delta_1$ and $\Delta_2$ are indicated. The inset shows the temperature dependences of the gaps $\Delta_1$ and $\Delta_2$.}
\label{Fig9}
\end{figure}

\begin{figure}[]
\includegraphics[width=8cm,angle=0]{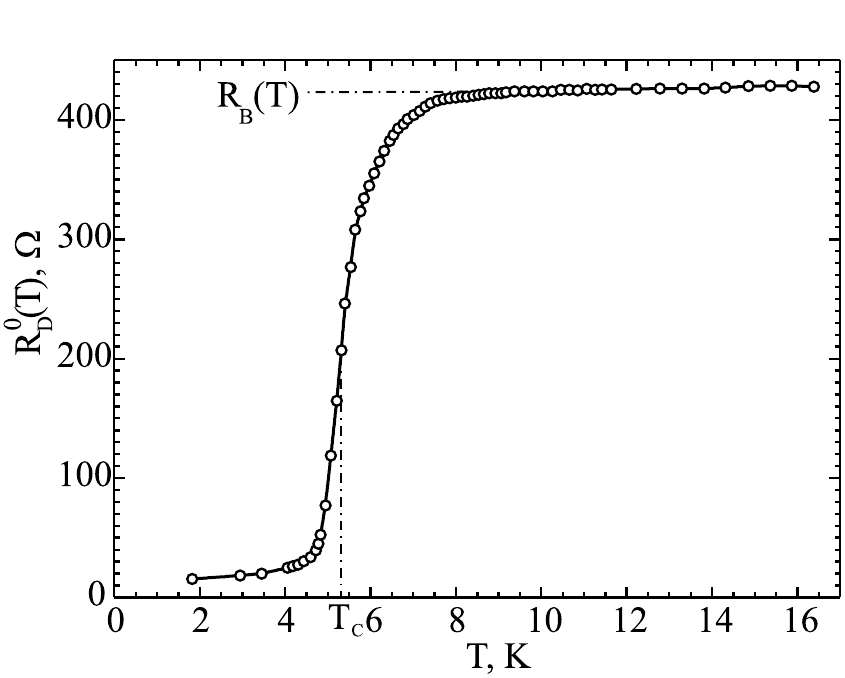}
\caption[]{Temperature dependence $R_{D}^{0}(T)$ of the differential resistance of the point contact $Cu$-SL3 (Fig.\ref{Fig9}) for zero bias voltage. The straight line $R_{B}(T)$ is the background linear variation of the point-contact resistance.}
\label{Fig10}
\end{figure}

\begin{figure}[]
\includegraphics[width=8cm,angle=0]{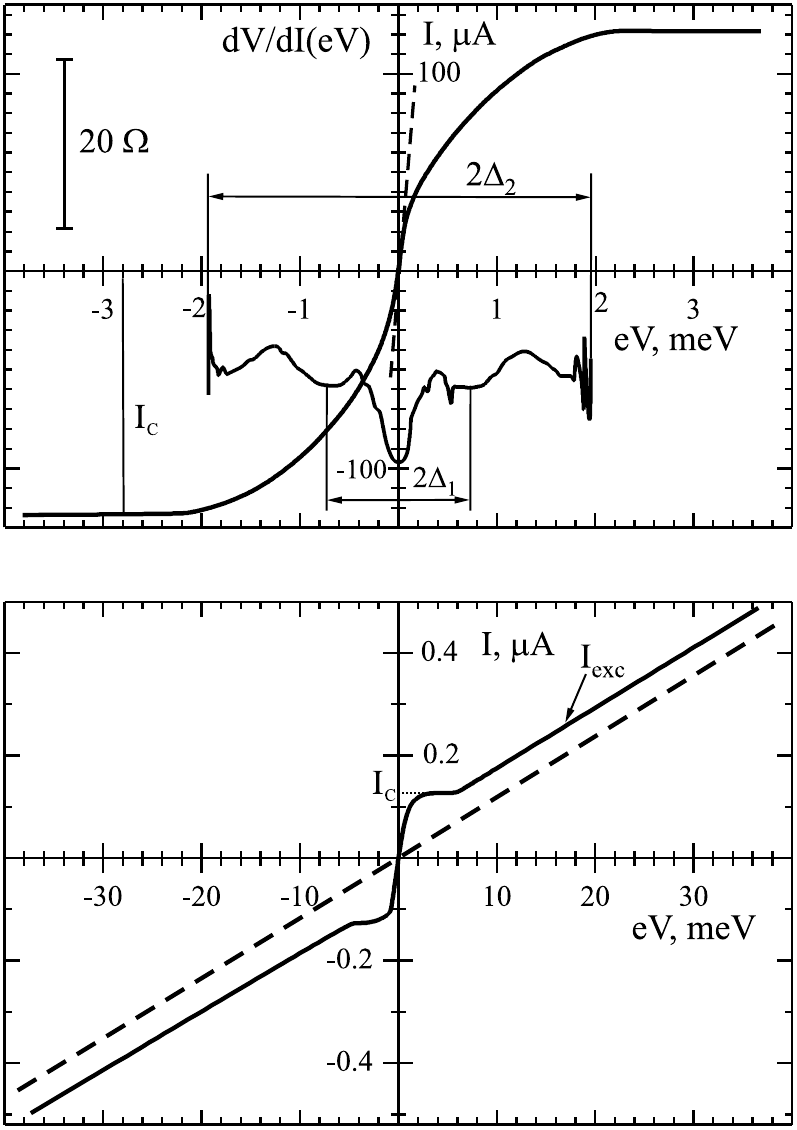}
\caption[]{IVC and its first derivative for the point contact $Cu$-SL4. The lower part of the figure shows the IVC over a wide range of bias voltages across the contact. The dashed lines help in determining the resistance $R_{D}^{0}(1.8~K)\approx 2\ \Omega$  for zero and $R_{D}^{V}(1.8~K)\approx 89\ \Omega$ for high bias voltages across the contact, excess current $I_{exc}\approx 62\ \mu A$ and "critical" current $I_{c}\approx 123\ \mu A$. The ways of determining the gaps $\Delta_1$, and $\Delta_2$ are shown.}
\label{Fig11}
\end{figure}

\begin{figure}[]
\includegraphics[width=8cm,angle=0]{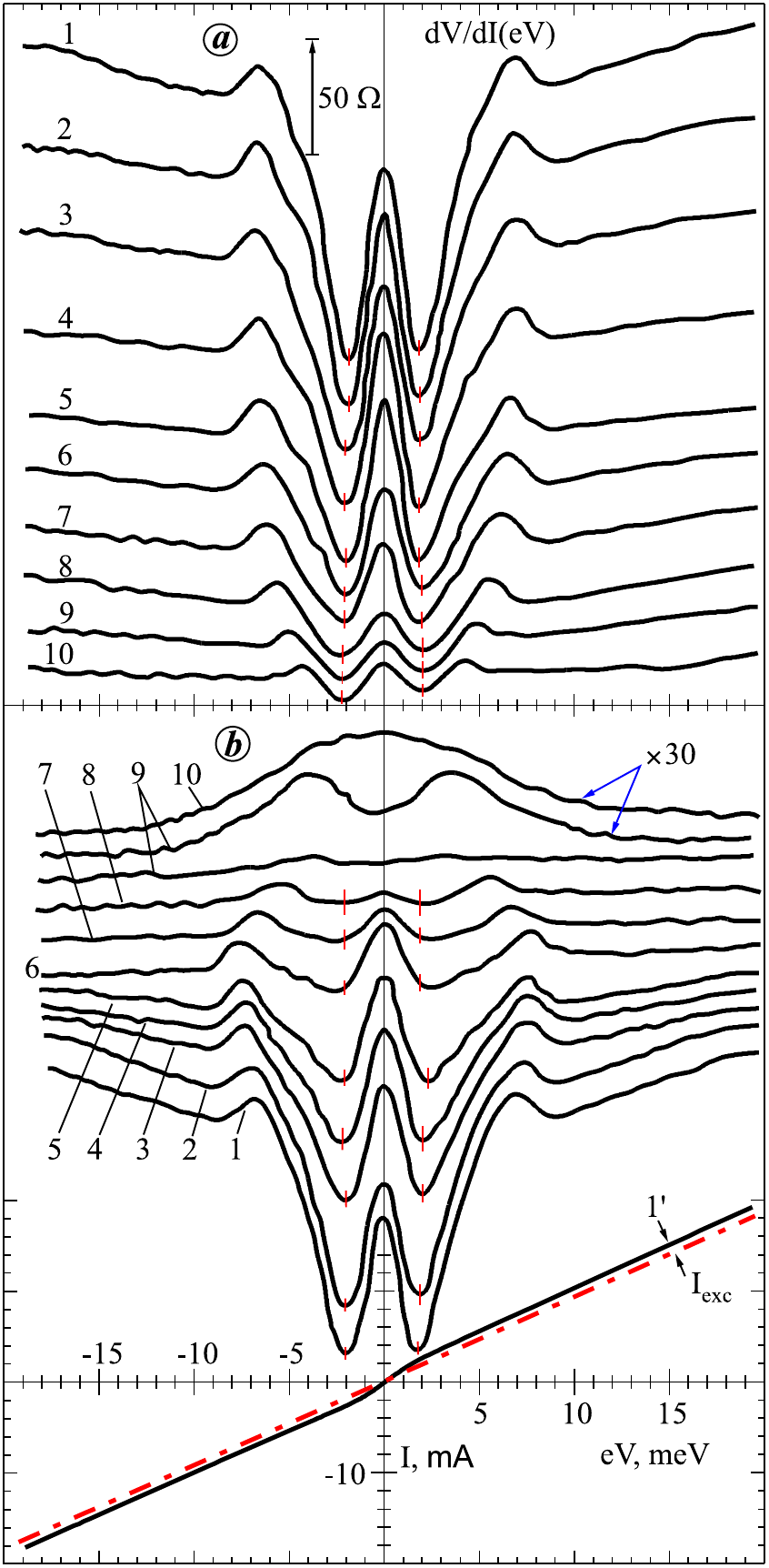}
\caption[]{IVC and its first derivatives for the point contact $Cu$-SL1; (a) dependence on magnetic field at $T=1.85~K:\ H$=0(1), 1.5(2), 2.6(3), 6.5(4), 10.8(5), 13.4(6), 18.2(7), 22.1(8), 26.3(9), and 30.2(10)~$kOe$; (b) temperature dependence for $H$=0:\ $T$=1.85(1,1'), 2.01(2), 2.56(3), 3.08(4), 3.5(5), 4(6), 4.55(7), 4.9(8), 5.5(9), and $6~K$(10). $R_{D}^{V}(1.85~K)=1050~\Omega$, and the dot-dash line is parallel to IVC for $eV\gg\Delta$.}
\label{Fig12}
\end{figure}

 lying in an interval where the differential resistance changes insignificantly (see Fig. \ref{Fig5}). This dependence cannot be attributed simply to the superconducting fluctuations, or even to the stronger effect of the emergence of superconducting clusters near the bridges. Investigations of the effect of superconducting clusters on the shape of IVC of point contacts \cite{31,32,33} showed that for large bias voltages across the contact, when the energy of the quasiparticles incident on a superconducting cluster is higher than $\Delta$, the electric field penetrates the cluster. If the size of the cluster is of the order of, or smaller than, the penetration depth of the electric field into the
 superconductor, the differential resistance coincides with that of the point contact in the normal state. The electric field penetration depth into the superconductor $l_{E}\sim l_{\varepsilon}$ ($l_{\varepsilon}$ is the mean inelastic scattering length for quasiparticles with energies $0<\varepsilon<eV$). For a layered superconductor like $NbSe_2$; which has values of $T_c$, $\varepsilon_\parallel$, and $\varepsilon_\perp$ of the same order as those for lead chalcogenide superlattices, we have $l_{E}\sim150~\AA$ \cite{34}. It can be assumed that the values of $l_{E}$ for super-lattices will also be similar. However, even if $l_{E}$ is considerably smaller than $d$ ($d$ is the contact diameter), for a sufficiently high bias voltage across the contact corresponding to a current density higher than the critical current density the electric field will penetrate the superconductor to a depth of the order of $d$ due to the formation of a system of phase slip centers, lines, or surfaces (PSS), and the differential resistance of the point contact becomes the same as in the $N$-state (see Figs.\ref{Fig1} and \ref{Fig2} in Ref. \cite{34}). The formation of the PSS is not always reflected on the IVC in the form of quasilinear segments on which the differential resistance changes abruptly. In Ref. \cite{35}, for example, the $dV/dI$ dependence of the $YBaCuO-Ag$ point contact under the action of a microwave field revealed oscillations in the region of a smooth increase in the excess current. Apparently, these oscillations are associated with the oscillations of the order parameter with the Josephson frequency on a PSS type weak bond in the contact region formed as a result of a considerable current injection. Similar oscillations were also observed for the $LaSrCuO$ ceramic in Ref. \cite{36}. In both cases, the IVC and the $dV/dI$ dependences were similar to those shown in Figs.\ref{Fig4} and \ref{Fig5} (curves 1, 1'). It should be observed that point contacts in which PSS are formed are not spectroscopic. Indeed, for the spectroscopy of quasiparticle excitations, two groups of electrons with energies differing by $eV$ must meet near the constriction. For the spectroscopy in the $S$-state, the chemical potential of Cooper pairs must remain unchanged right up to the constriction plane (see Fig. \ref{Fig3} in Ref. \cite{37}). In the presence of a PSS, however, the chemical potential of the pairs suffers a discontinuity. This does not mean that the formation of a PSS is inevitable for all point contacts with $d\gg l_\varepsilon$. For example, the existence of normal inclusions in the current flow region near the contact which have a higher quasiparticle conductivity will shunt the $S$-channel and prevent the attainment of critical current density, and hence the formation of PSS, in it. Apparently, this circumstance made it possible to register the EPI spectra in $LaSrCuO$ and $YBaCuO$ \cite{36,37,38}. Since point contacts have been used in our work to register the EPI spectra in SL (see below), it can be stated that PSS are not formed at least in these point contacts.

In the case of superlattices (Fig.\ref{Fig8}), the differential resistance of point contacts for large bias voltages (the curves were recorded up to 100 $mV$) begins to decrease even in the region of superconducting fluctuations, attaining its minimum value after a transition of the superlattice to the superconducting state (at $T\sim {0.7\ T_c}$, when SC is localized in $PbTe$ layers). In many cases, this decrease is quite significant (many-fold; see, e.g., the IVC in Fig.\ref{Fig9}). The decrease in $R^V_D(T)$ may be due an increase in the normal (quasiparticle) conductivity of the material in the vicinity of a bridge. The temperature dependence of the variation of this conductivity was found to be the same as for fluctuational superconductivity in normal superconductors, and apparently also reflects the geometrical dimensionality of regions in which the conductivity increases. The normalized excess conductivity \[\frac{{\sigma }'(T)}{\sigma (9\ K)}=\frac{R_{D}^{V}(9\ K)-R_{D}^{V}(T)}{R_{D}^{V}(T)}\] constructed in logarithmic coordinates (see triangles in Fig.\ref{Fig2}) as a function of the reduced temperature $\tau=(T-T_{c})/T_c$ in the same way as for the superconducting fluctuations, shows that the dependence $\sigma'(T)$ (see Table \ref{tab2}) corresponds to three-, two-, and zero-dimensional regions over quite a wide temperature interval. As in the case of an analogous construction for superconducting fluctuations, the mid-point of the superconducting transition ($T_{c}=3.5~K$) was taken as the critical temperature. The fact that this construction containing the superconducting transition temperature $T_c$ gives a temperature dependence corresponding to the Aslamazov-Larkin theory indicates that this phenomenon is connected with superconductivity. it was mentioned in the introduction that electron states with a binding energy of the order of $\varepsilon_F$ between the carriers and the heteroboundary are localized near the latter. The conductivity along the heteroboundary in the $N$-state is low due to a large disorder caused by the existence of the dislocation network (the system is close to Anderson localization). Since the superconductivity is due just to the presence of the dislocation network, the superconducting transition apparently entails a radical rearrangement of the electron spectrum accompanied by an insulator-metal transition. This transition can be seen in Figs.\ref{Fig5}-\ref{Fig7}, \ref{Fig9}, and \ref{Fig12}. While the first IVC derivative in the $S$-state is of metallic type, i.e., the resistance increases with increasing bias voltage across the contact, the $N$-state is characterized by a semiconductor-type dependence (or $R_{N}\sim$const in the entire range of bias voltages across the contact). A similar behavior, which was observed earlier during the investigation of $LaSrCuO$ \cite{36} and $YBaCuO$ (Ref. \cite{39}), is also apparently characteristic of HTS materials.

In the Aslamazov-Larkin fluctuational conductivity model \cite{18}, the increase in conductivity is associated with an increase in the number of Cooper pairs. Similarly, the normal quasiparticle conductivity is proportional to the carrier concentration. Hence it can be assumed that the formation of Cooper pairs is accompanied by the delocalization of a proportional number of electron states at the heteroboundary. Apparently, this is why the temperature dependence of conductivity for large bias voltages follows the temperature dependence of the fluctuational superconductivity.

Thus, high-conductivity regions are first formed in the vicinity of the dislocation network sites, and then merge with the high-conductivity heteroboundary. The temperature at which a transition from $2D-$ to 3$D$-regions takes place in this model will generally depend on the diameter of the point contact and its separation from the heteroboundary. The same parameters determined the relative decrease in $R_{D}^{V}(T)$. In contrast to the measurements of $\sigma (T)$ for superlattices as a whole (squares in Fig.\ref{Fig3}) which show an exponent $k$=3/4 over a sufficiently wide temperature range corresponding to the case intermediate between two- and three-dimensional regions, the excess conductivity of a point contact clearly marks out the $0D-2D-3D$ regions (triangles in Fig.\ref{Fig3}). Apparently, this is due to the fact that an increase in conductivity in the former case is associated with the formation of high-conductivity regions as well as the superconducting fluctuations, while the temperature-dependent part of the conductivity for a point contact is mainly associated with electrons delocalized from the MD network region.

By measuring the temperature dependences $R_{D}^{V}(T)$ and $R_{D}^{0}(T)$ for the same point contact, one could expect to separate in pure form the excess conductivity associated only with superconducting fluctuations. Unfortunately, on account of a considerable temperature-dependent tunneling and a nonzero order parameter ("energy gap"?) in the fluctuation region (see curves 10-26 in Figs.\ref{Fig5} and \ref{Fig8}), the plot of $\sigma'(T)$ using the $R_{D}^{0}(T)$ dependence for the given contact does not fit in any theoretical dependence. It was mentioned above that the maximum difference between the values of $R_{D}^{V}(T)$ in $N$- and $S$-states must be observed for a point contact formed directly at the heteroboundary. Apparently, Fig.\ref{Fig9} corresponds to the characteristics of such a contact. In the $S$-state at $T=4.2~K$, $R_{D}^{0}(4.2)\simeq96~\Omega$ in the bias voltage range 10-20 $meV$, where the differential resistance does not vary significantly. On the other hand, in the $N$-state at $T=10~K$, $R_{N}\simeq419\  \Omega\ (R_{N}/R_{N}^{V})\ (9.2~K)\simeq 4.4$  over the entire range of bias voltages. In view of a very low tunneling capacity of the contact as indicated by the absence of a maximum in the differential resistance of the point contact for zero bias voltage, and also the temperature-independence of the tunneling due to a close proximity of the contact to the heteroboundary, the temperature dependence of the excess conductivity for zero bias voltage was used to determine the size of the regions in which the conductivity increases for $T>T_c$. Figure \ref{Fig10} shows the temperature dependence $R_{D}^{0}(T)$ for a given point contact used for determining the dimensionality of superconducting fluctuations (+ in Fig\ref{Fig3}). For this SL3, like for the preceding SL1, the fluctuation range extends to $7~K$ in spite of the fact that the critical temperature is much higher in this case: $T_{c}=5.3~K$. The fluctuations are found to be three-dimensional in the narrow temperature interval in the vicinity of $T_{c} (5.4<T<5.5~K)$. This is followed as in the measurement of the normal conductivity SL1, by a region with $k=3/4$ which is replaced by a narrow transition region with $k=1$, ensuring a transition to the region of zero-dimensional fluctuations (see Table \ref{tab2}). For the point contact under investigation, the IVC for large bias voltages is highly nonlinear on account of inelastic scattering of electrons. As a result, the conductivities of the point contact associated with normal quasiparticles are different for $V=0$ and $V\gg \Delta_{0}$. Hence the contribution of fluctuation pairs alone to the conductivity cannot be isolated by knowing the temperature dependences $R_{D}^{0}(T)$ and $R_{D}^{V}(T)$.

Summing up, it can be stated that the formation of fluctuational Cooper pairs is apparently accompanied by the delocalization of electron states in the heteroboundary region. Thus, the change in the differential resistance for large bias voltages is associated with a change in the geometry of current flow near the point contact. The flow is more three- dimensional in the $N$-state, two-dimensional in the $S$-state, and takes place predominantly along the highly conducting heteroboundary. Delocalization of electrons from the vicinity of structural elements responsible for SC obviously explains the change in the nature of the $R_{D}^{V}(T)$ dependence during SC transition. In Ref. \cite{36}, this phenomenon was conditionally called the insulator-metal transition.

Note that the temperature measurements of $R_{D}^{0}(T)$ for the point contact $Ag-YBa_{2}Cu_{3}O_{7-\delta}$ (Ref. \cite{39}) also reveal regions of 3$D$- and 2$D$-fluctuations. The region of zero-dimensional fluctuations is quite narrow and blurred. The beginning of the 2$D$- and 3$D$-regions (with decreasing temperature) corresponds approximately to $\tau=1, {(t/2t')}^2$. The small width of the 0$D$-fluctuation region in this work is associated with an insufficient accuracy of measurements, as well as the fact that it is hard to keep the point contact unchanged over such a wide temperature range. A study of the shape of $dV/dI$ curves shows that symptoms of metal-insulator transition in $YBaCuO$ point contacts appear for $T<220~K$, first in the form of a flattening and then, upon a further decrease in temperature, in the form of a minimum in the vicinity of $V=0$ on the bell-shaped semiconductor-type $dV/dI$ curve. This leads to the assumption that the 0$D$-fluctuation region extends to $220~K$ in $YBaCuO$ (the value $\xi=t'/2$ corresponds to $T=210~K$ for $T_c = 90~K$). It is quite significant that $dV/dI$ has a semiconductor behavior for $T\geq 220~K$, i.e., the differential resistance decreases with decreasing bias voltage across the contact. However, $R_{D}^{0}(T)$ increases with temperature, i.e., has a metallic dependence. On the whole, the dependence $R_{D}^{0}(T)$ for a point contact is similar to the $R(T)$ dependence measured for a bulk sample by the conventional resistive technique. Thus, the fluctuation range and the limits of the variation of their dimensionality, determined by using the $R_{D}^{0}(T)$ dependence for a point contact and by conventional resistive measurements for a bulk sample, do not coincide. Apparently, this is connected with the existence of a high-$T_c$ phase in the vicinity of the contact.

\subsection{Estimate of the characteristic contact size. Critical density and excess current in point contacts}
Let us estimate the diameters of all the contacts described above. For this purpose, we shall use Maxwell's formula $d\approx \rho/2R$ neglecting the contribution of the copper bank to the resistance. The values of $\rho$ for superlattices are presented in Table \ref{tab1} (it should be remembered that these are the values of $\rho$ averaged over the thickness of SL; the real value of $\rho$ of a material in the vicinity of a constriction is determined in each specific case by the layer in which the point contact is formed, as well as by the separation between the heteroboundary and the bridge). For $R$, we use the normal state resistance $R_N$, characterizing the maximum three-dimensionality of the current flow. For contacts in which the value of $R_N$ has not been measured, the diameter was estimated by using the available data $R_{D}^{V}(T)$ (the same formula was used for computing the value of $d$). The obtained values are compiled in Table \ref{tab3}.
\begin{table}[]
\caption[]{}
\begin{tabular}{|c|c|c|c|c|cc|c|} \hline
\multicolumn{1}{|c|}{\centering \multirow{2}{*}{Figure}}& \multicolumn{1}{|c|}{\centering \multirow{2}{*}{$R_{N}, \Omega $}} & \multicolumn{1}{|c|}{\centering \multirow{2}{*}{
$d_{N}$, \AA}} & \multicolumn{2}{|c|}{$\Delta(1.8~K)$} & \multicolumn{2}{|c|}{\centering \multirow{2}{*} {$2{\Delta}/kT_{c}$}} & Superlattice   \\
& & &   \multicolumn{2}{|c|}{ $meV$}& &\  & number \\ \hline
4, 13a, 14a & 255\footnote{$R_{D}^{V}(1.8~K), \Omega$} & 306 & \multicolumn{2}{|c|} {2.1}&\multicolumn{2}{|c|} {12.5}& 1\\
5&1180&70&\multicolumn{2}{|c|}{2.7} &\multicolumn{2}{|c|}{16}&1\\
6&190\footnote{$R_{D}^{0}(1.75~K), \Omega$}&212&\multicolumn{2}{|c|}{1.7} &\multicolumn{2}{|c|}{10}&2\\
7&170&233&\multicolumn{2}{|c|}{2.35} &\multicolumn{2}{|c|}{14}&2\\ \cline{4-7}
9&420&160&0.8&2.4&3.5 \vline &10.5&3\\
11&$90^{a}$&310&0.75&2&3.2 \vline&8.4&4\\ \cline{4-7}
12&$1050^{a}$&75&\multicolumn{2}{|c|}{2} &\multicolumn{2}{|c|}{7.7}&1\\
13b, 14b&$460^a$&60&\multicolumn{2}{|c|}{-} &\multicolumn{2}{|c|}{-}&4\\ \hline
\end{tabular}
\label{tab3}
\end{table}
 The accuracy of these estimates is different for different contacts. For example, the estimate is found to be too low for the contact in Fig.\ref{Fig7}, if we consider the strong tunneling in this contact which was not taken into account during calculations. The estimate is quite good for the contact shown in Fig.\ref{Fig11} (obtained by the method of pulsed electric short-circuiting). For this contact, the inequality $d\gg \xi_{\perp}$ is satisfied at low temperature. One of the quality criteria for such contacts is close values of contact diameters obtained in the superconducting state from the zero-bias resistance by using Sharvin's formula, and in the normal state by using Maxwell's formula \cite{36}.

The  above condition assumes that the superconducting phase is in direct contact with the interface and covers the entire contact region. Since the transformation of quasiparticles into pairs occurs for $eV<\Delta$ at distances $\xi\ll d$, the superconducting bank does not contribute to the point contact resistance. In the case of ballistic mode of electron transit in
the copper bank, the coincidence of Fermi parameters of the semiconductor and copper, and the absence of additional scatterers at the boundary, the point contact resistance taking into account the Andreev reflection would be half the value of a copper homojunction of the same diameter. Thus, neglecting the contribution of the copper bank in the $N$-state to the resistance, we obtain from this model \[{{d}_{s}}={{\left[ \frac{8{{\left( \rho l \right)}_{Cu}}}{3\pi R_{D}^{0}(T)} \right]}^{{1}/{2}\;}}=\frac{212}{\sqrt{R_{D}^{0}(T)(\Omega )}}\ (\AA);\] \[{{d}_{N}}=\frac{{{\rho }_{m}}}{2R_{D}^{V}(T)}=\frac{278\cdot {{10}^{2}}}{R_{D}^{V}(T)(\Omega )}\ (\AA).\]

Here, we have used the values $(\rho l)_{Cu}\simeq 0.53\cdot 10^{-11}\Omega\cdot cm^2$ (Ref. \cite{40}) and $\rho_{m}$ was taken from Table \ref{tab3} for SL4. Such an idealized situation for the contact shown in Fig.\ref{Fig11} is realized only in order of magnitude, i.e., $d_{s}\approx 150~\AA,  d_{N}\approx 310~\AA$. This is caused by the barrier reflection of carriers at the copper-semiconductor boundary due to different values of the Fermi parameters and $l_i$ at the banks ($l_i$ is the elastic mean free path of electrons). In other words, a slight tunneling occurs but is not taken into account while obtaining the estimates. Hence the point contact resistance for $eV<\Delta$ is found to be higher than the resistance of a copper homojunction of the same diameter. Moreover, the current flow from copper side may not be ballistic, but intermediate, or even diffusive. Hence the estimate of $d_s$ gives only the lower limit of the contact diameter.

Knowing the critical current corresponding to the breakdown current ($I_{c}\approx 123~\mu A$) on the IVC and the diameter of the point contact, we can similarly \cite{41} estimate the critical current density $j_{c}\approx(1.6-7)\cdot 10^{7}A/cm^2$ in the superlattice. This value is the lower estimate on account of the spatial inhomogeneity of the superconducting properties of the superlattice (the maximum values of parameters are attained at the heteroboundary). Conventional resistive measurements give the value of $j_c$ which is four orders of magnitude lower \cite{5}. This is due to weak couplings between the mosaic blocks.

Note that the excess current for our contacts is much higher than the value of the current obtained from the theory of $S-c-N$ contacts of small diameter ($d\ll\xi$) \cite{42}, especially if we take into account the considerable tunneling for a number of contacts which, in accordance with the theory \cite{42}, must lead to a negligible value of the excess current. The mechanism of formation of IVC with excess current for SL-based point contacts is apparently similar to the mechanism proposed in Ref. \cite{43} by Iwanyshyn and Smith, but is not thermal mechanism. It was mentioned in the preceding section that the formation of Cooper pairs in an SL is accompanied by the delocalization of a proportional number of quasiparticle excitations from the heteroboundary region. This delocalization of carriers considerably increases the level of quasiparticle conductivity as compared to the SL conductivity in the normal state. For $eV>2\Delta$, Cooper pairs in the contact region are destroyed as a result of collisions between electrons. In accordance with the model proposed above, this causes a decrease in the quasiparticle conductivity, i.e., the region with a low quasiparticle conductivity extends from the plane of the hole (The model proposed by Iwanyshyn and Smith considers the displacement of the $N-S$ boundary from the plane of the hole associated with the Joule heating of the contact region.) This leads to a large excess current (up to $10\Delta/R_{D}^{V} (1.7~K$) if we measure $I_{exc}$ as the separation between the IVC and the straight line parallel to it and passing through the origin for $eV\gg\Delta$). Apparently, an analogous mechanism of formation of the excess current also prevails for the $YBaCuO$ based point contacts.

\subsection{Temperature and magnetic field dependence of the energy gap}
For small-diameter $S-c-N$ point contacts ($d\ll\xi$) containing a tunnel component of the current, the superconducting energy gap appears on the $dV/dI$ dependence in the form of peaks arranged symmetrically relative to the ordinate axis at energies $eV\approx\pm \Delta$ \cite{42}. In contacts with an extremely low tunneling, the excess current is of the order of $\Delta/eR$ and decreases rapidly upon an increase in tunneling. Note that even for normal $S-I-N$ tunnel contacts with different tunneling capacities, the theory predicts an analogous manifestation of the energy gap \cite{44}. However, this theory does not take into account factors like current flow in the vicinity of the contact, the influence of nonequilibrium effects associated with the relaxation of the imbalance between the electron and hole branches of quasiparticles in a superconductor, as well as the possibility of formation of phase slip surfaces in the contact region.

For the contacts investigated by us, the current concentration cannot be disregarded, and the intermediate relation between $d$ and $\xi$ is satisfied for most of them: $d<\xi_\perp$ in the vicinity of $T_c$, while $d>\xi_\perp$ at lower temperatures. There is no theory explaining the mechanism of formation of gap peculiarities on the IVC of such point contacts. The situation becomes even more complicated due to a spatial inhomogeneity of superiattices with $\xi_\perp$ smaller than the lattice period at low temperatures. Nevertheless, we shall associate the emergence of such minima on the $dV/dI(V)$ dependence with the superconducting energy gap $\Delta(T, H)$.

In SL1 and SL2, superconductivity (and the gap) are induced in $PbS$ for $T\ll T_c$ in the immediate vicinity of the heteroboundary. For contacts with a considerable tunneling without the excess current, the bridge is formed in the bulk of the $PbS$ layer which is depleted of carriers and plays the role of a tunnel barrier. Since the order parameter has its maximum value at the heteroboundary and rapidly decreases to zero in the bulk of $PbS$, it seems interesting to find which gap is registered by the point contact. This question has been
answered in Ref. \cite{45}, where the point contact is created between a silver needle and a silver film deposited on lead surface without a barrier. For a silver film of small thickness, the contact registers a gap induced in the vicinity of the constriction due to the proximity effect. With increasing thickness of the silver film, the induced gap width decreases. For a large thickness of the silver film, when the width of the gap induced in the silver contact was equal to zero, the point contact registered the unperturbed gap corresponding to lead. The jump $\Delta$ at the boundary between lead and silver is the singularity that makes this possible. Similarly, in our case also, the contact apparently registers the unperturbed gap in $PbTe$ in the vicinity of the heteroboundary (see Figs.\ref{Fig6} and \ref{Fig7}).

It was mentioned above that unlike SL1 and SL2 in which superconductivity is quasi-two-dimensional at low temperatures, SL3 and SL4 are three-dimensional anisotropic systems. For point contacts based on these superlattices and shown in Figs.\ref{Fig9} and \ref{Fig11}, two gaps are observed. This is probably due to the fact that the point contacts are located at the heteroboundary and behave like two point contacts connected in parallel: $Cu-PbS$ and $Cu-PbTe$. The smaller gap is induced in $PbS$ and the larger in $PbTe$. The values $\Delta_1$ and $\Delta_2$ of the gaps and their ratios $2\Delta / kT_c$ are presented in Table \ref{tab3}. For the contact shown in Fig.\ref{Fig9}, the temperature dependence of $\Delta_1$ and $\Delta_2$ was followed in the interval $1.86-3.7~K$. At low temperatures, $\Delta_1$ and $\Delta_2$ were compared with the minima on $dV/dI$. With increasing temperature, these singularities are rapidly blurred. Hence the gaps were traced by recording the second derivatives on which the values of $\Delta_1$ and $\Delta_2$ were associated with the minima on $d^{2}V/dI^2$ by joining of gap dependences in low and high temperature regions. By way of an example, Fig.\ref{Fig9} shows the two curves $d^{2}V/dI^2$
for $T=3$ and $3.7~K$. The curves are completely blurred at $T>3.7~K$, and only one structureless peak is left on the $d^{2}V/dI^2$ curve. Note the similarity between the curves 7 and 8 in Fig.\ref{Fig9} and the curves 7 and 8 in Fig.\ref{Fig4}; both these cases correspond to gapless superconductivity. The $\Delta_{1}(T)$ and $\Delta_{2}(T)$ dependences constructed for this contact are shown in Fig.\ref{Fig9} (lower inset) and differ from the BCS dependence in a more rapid decrease in the gaps at low temperatures.

It is remarkable that for SL3 and SL4 (Figs.\ref{Fig9} and \ref{Fig11}), the ratio of the gaps $\Delta_{2}/\Delta_{1}\approx 3$ for $1.8~K$. For SL3, the larger gap appears at $T=0.75~T_c$ i.e., when the order parameter is localized in the $PbTe$ layer, while smaller gap appears approximately at $T\sim 0.7\ T_c$. The value of $2\Delta_{1}(1.8~K)/kT_c$ for both SL is close to the value 3.54 in the BCS theory, while the larger gap has a value $2\Delta_{2}(1.8~K)/kT_{c}\sim 10$ characteristic of HTS materials. Two gaps with quite identical temperature dependences (and gapless conductivity in a wide temperature range) were discovered in $La_{1.8}Sr_{0.2}CuO_4$ \cite{36}, but the smaller gap was found to vanish at $T=(T_{c}/2)$ and the larger one at $T\sim 0.73~T_c$, while the gaps at $T\ll T_c$ differ by a factor of about two. Two tunnel singularities of the gap were also observed for thin epitaxial films of $YBa_{2}Cu_{3}O_7$ (Ref. \cite{46}) on the curve $dI/dV$ ($\pm 16$ and $\pm 30~mV$ at $T=4.2~K$). The separation between peaks on $dI/dV$ remained practically unchanged up to $T\rightarrow T_c$. The gap structure
disappeared at $T\rightarrow T_c$ as a result of a continuous "softening" and not due to a displacement of the peaks towards lower energies.

The problem concerning the reasons behind the emergence of multigap structure is not trivial, and there are several ways of explaining this phenomenon (see, e.g., Refs. \cite{23} and \cite{47}). In any case, the condition $l>\xi_0$ must be satisfied for the observation of several gaps \cite{47}. From this point of view, the condition is more easily satisfied in an SL with a high $T_c$, and this is probably the reason why two gaps were observed for such SL only.

In contacts with direct conductivity, a high current density leads to the emergence of gapless superconductivity in the vicinity of $T_c$ and hence to a much more rapid decrease in the gap with temperature than predicted by the BCS theory. In tunnel contacts, the current density effects are not significant and, according to Ref. \cite{46} as well as our results for contacts with a tunnel barrier, the gap is practically independent of temperature right up to $T_c$ (see Figs.\ref{Fig6} and \ref{Fig8}).

Figures \ref{Fig4} and \ref{Fig7} show the $dV/dI$ characteristics of point contacts based on SL1 and SL2, recorded in different magnetic fields oriented parallel to the films, and the dependences $\Delta(H)$ for these curves. The values of $\Delta(0)$ and the ratio $2\Delta/kT_c$ are presented in Table \ref{tab3}. The gap vanishes, i.e., gapless superconductivity sets in, for the first contact in fields stronger than $15~ kOe$, and for the second contact in fields stronger than $18~kOe$. Note that such a dependence of $\Delta(H)$ is characteristic of thin films.

Figures \ref{Fig5} and \ref{Fig6} show the families of curves $dV/dI(eV)$ recorded at different temperatures, while Fig.\ref{Fig8} shows the temperature dependence of the gaps for contacts based on SL1 and SL2. Like the coherence length $\xi_{\perp}(T)$ (Fig.\ref{Fig2}), the gaps show a nonmonotonic temperature dependence $\Delta(T)$. Both dependences $\Delta(T)$ coincide with the BCS dependence
$\Delta(T)/\Delta(0)$ only in a narrow temperature interval ($2.5<T<3.2~K$). At lower temperatures, the average gap in the contact region decreases since superconductivity is drawn towards the heteroboundaries. The gap and zero-dimensional fluctuational conductivity appear simultaneously. The transition to quasi- two-dimensional fluctuational conductivity corresponds to the onset of the descending region on $\Delta(T)$ curve.

In the fluctuation region, the $dV/dI$ dependences have minima identical to the gap minima whose position on the energy axis is practically independent of temperature. It can be assume that these minima are connected with superconductivity localized at the heteroboundary, while the independence of the position of gap minima on the energy axis from temperature or magnetic field can be associated with the specific nature of this phase as, for example, in Kulik's theory where the gap for a pair may be independent of temperature \cite{23}. In "pure" form, such superconductivity was observed in a point contact formed apparently at the second heteroboundary of SL1 (Fig.\ref{Fig12}). With increasing magnetic field (\emph{a}) or temperature (\emph{b}), the depth of the gap minima decreases although their position on the energy axis remains unchanged.

An identical behavior of the gap was also observed in HTS materials. In Ref. \cite{39}, for example, the $dV/dI$ dependence for a point contact based on $YBa_{2}Cu_{3}O_{7-\delta}$ single crystal at $T>T_c$ showed broad minima near $\pm eV\approx 100~meV$ in the region of SC fluctuations, the position of these minima being independent of temperature (see Ref. \cite{38} also). For the $Bi_{2}(Sr_{0.6}Ca_{0.4})_{3}Cu_{2}O_x$ polycrystal \cite{48}, the separation between $dI/dV$ peaks depends weakly on temperature in the case of tunneling through the amorphous $a$-Si layer. With increasing temperature, the peaks gradually disappeared.
\subsection{Electron-phonon interaction in superlattices}
In order to understand the nature of superconductivity in superlattices, it seems interesting to study the electron-phonon interaction (EPI) spectrum. It can be assumed that this spectrum will not be a superposition of the EPI spectra of $PbTe$ and $PbS$. A significant role in the formation of the spectrum must be played by the phase boundaries on which the square MD network with a period 52~\AA\  is located, the deformation potential of dislocations extends into each layer to a depth of the order of the dislocation structure period, while the dislocations themselves are arranged one over the other to form a simple tetragonal lattice. Such a lattice can be presented as an ideal crystal with its own vibrational frequencies.
\begin{figure}[]
\includegraphics[width=8cm,angle=0]{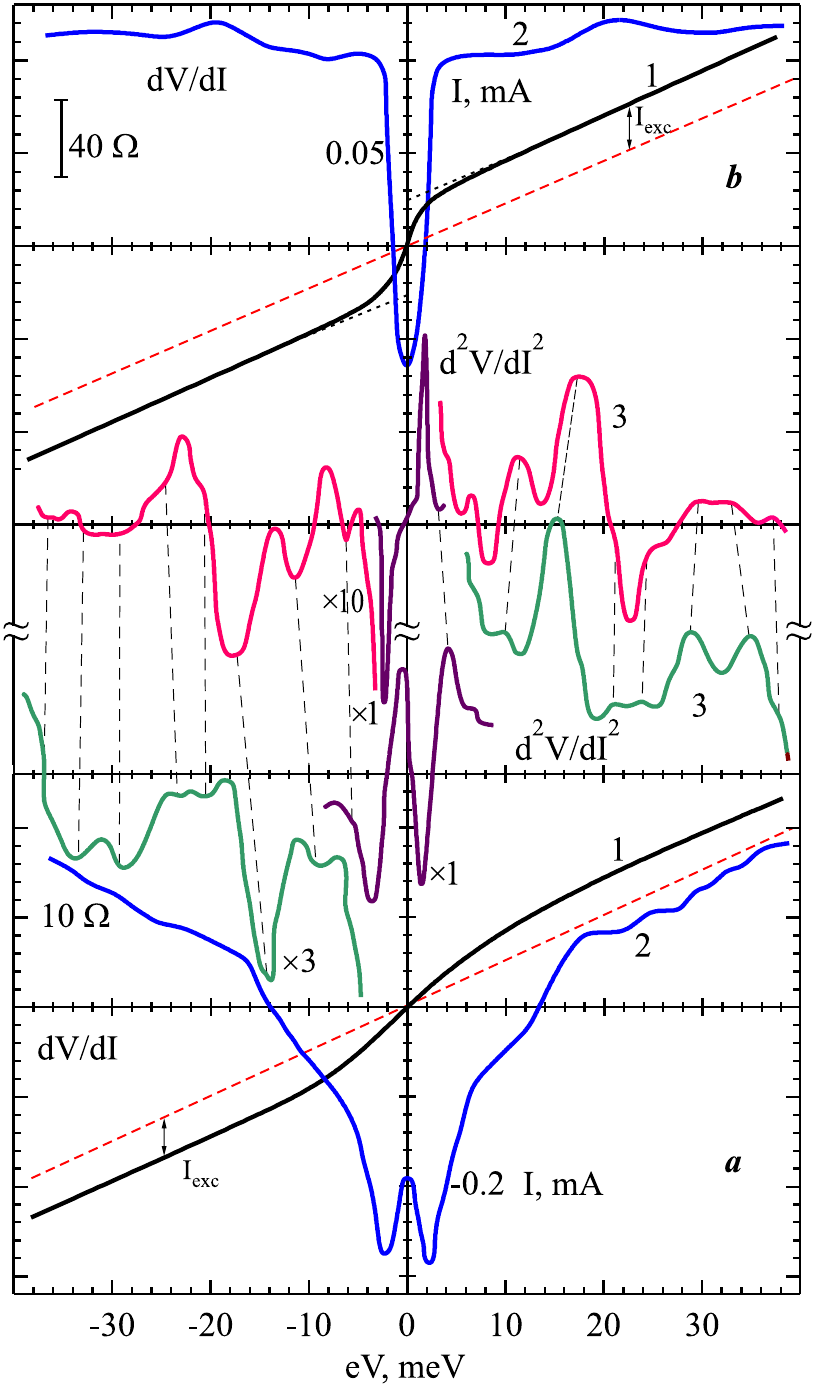}
\caption[]{IVC (1) and their first (2) and second (3) derivatives for the point contacts $Cu$-SL, $H$=0: (a) the same contact as in Fig. 4, $T=1.85~K$, $R_{D}^{V}(1.85~K)=225~\Omega$; (b) SL4-$Cu$, $T=1.8~K$,  $R_{D}^{0}(1.8~K)=60~\Omega$, $R_{D}^{V}(1.8~K)=460~\Omega$.}
\label{Fig13}
\end{figure}

\begin{figure}[]
\includegraphics[width=8cm,angle=0]{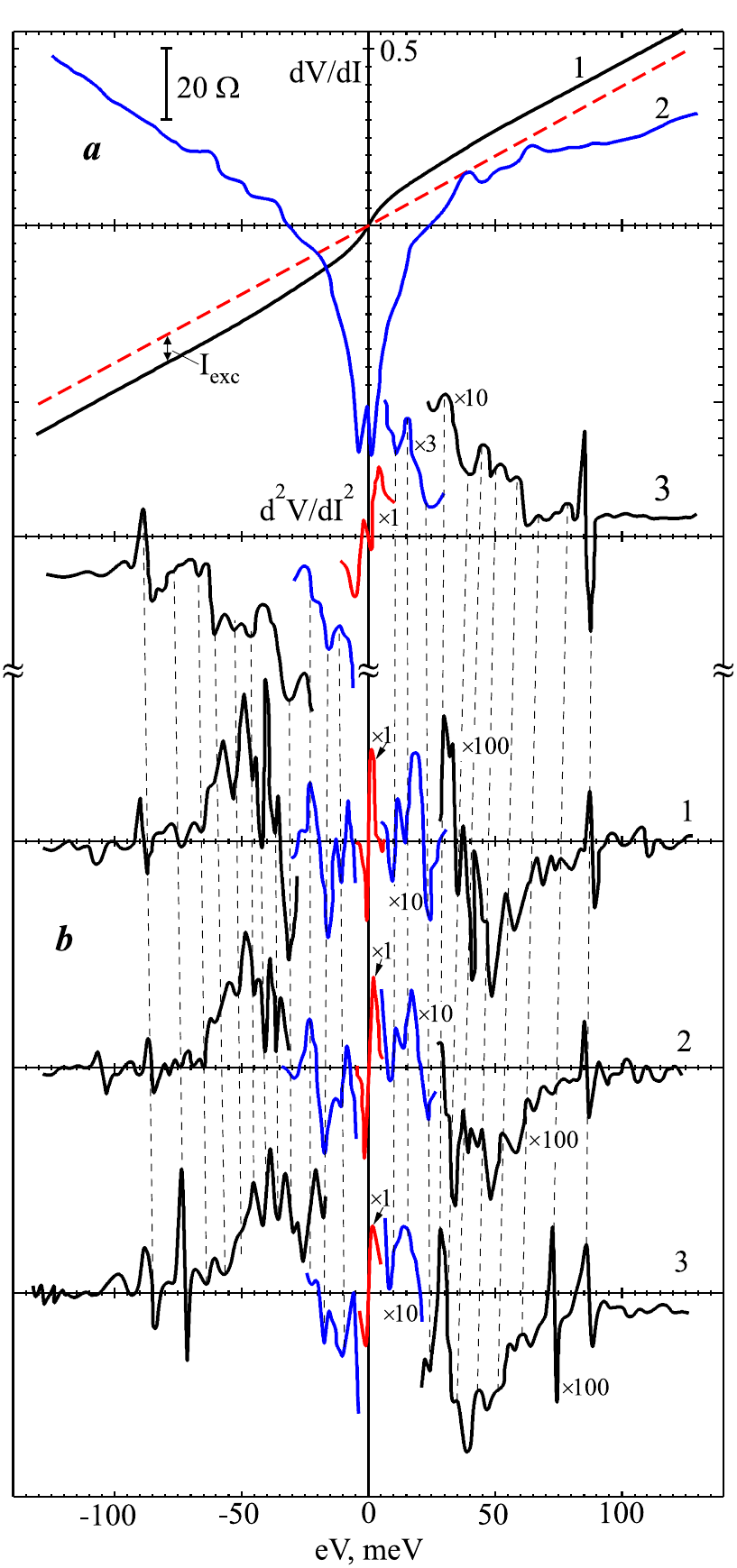}
\caption[]{The same as in Fig.13 in a wide range of bias voltages: (a) $T=1.86~K$(1,2,3); (b) $T$=1.85(1), 2.38(2), and $3.25(3)~K$. The dashed lines are drawn to facilitate the tracing of the phonon singularities position in the spectrum upon a change in temperature.}
\label{Fig14}
\end{figure}
Investigations were carried out by using SL1 in which superconductivity in quasi-two-dimensional with a weak Josephson coupling between layers (Figs.\ref{Fig13}\emph{a} and \ref{Fig14}\emph{a}) and SL4 with a three-dimensional anisotropic superconductivity (Fig. \ref{Fig13}\emph{b}, \ref{Fig14}\emph{b}). A possible mechanism for the emergence of singularities on EPI spectra in superconducting point contacts was proposed in Ref. \cite{37} and is based on reabsorption by Cooper pairs of phonons with low group velocities ($\partial\omega /\partial q=0$) which slowly leave the contact region and cause a local decrease in the energy gap. The excess current in such point contacts is due to inelastic processes of Andreev reflection near the $N-S$ boundary and is proportional to $\Delta$. For bias voltages corresponding to characteristic phonon energies, the excess current also decreases together with $\Delta$, which is reflected in the form of peaks on the IVC derivatives. The manifestation of phonon singularities on IVC derivatives in SL is apparently associated with the specific nature of the formation of excess current in these point contacts. Since the number of quasiparticle carriers delocalized from the regions in the vicinity of heteroboundaries is proportional to the concentration of cooper pairs, reabsorption of phonons with $\partial\omega /\partial q=0$ by Cooper pairs results in a localization of carriers, and hence to a decrease in the quasiparticle conductivity. Apparently, a similar mechanism is also responsible for the manifestation of phonon singularities in HTS \cite{36,38}.

Thus, the peaks observed on the second derivatives of IVC correspond to the Van Hove singularities of the functions $F(\omega)$ of the phonon densities of states in $PbTe$ and $PbS$, as well as to the local and quasilocal vibrational modes associated with the existence of MD networks in SL. For both SL, the energies corresponding to the positions of these peaks were found to be close and restricted to a value $90~meV$. These energy values are presented in Table \ref{tab4} for $T=1.8~K$. The constant value of the excess current over the entire range of bias voltages, the existence of a welldefined limiting frequency of the spectrum, and a practically constant energy position of these singularities on the spectra upon a change in temperature over a wide range (for SL4) confirm their spectral origin and rule out the possibility of appearance of parasitic peaks in spectra associated with the degradation of superconductivity.
\begin{table}[]
\caption[]{}
\footnotesize
\begin{tabular} {|c|c|c|c|c|c|c|c|c|c|c|c|c|c|c|c|c|c|c|} \hline
SL,& \multicolumn{18}{|c|} {\centering \multirow{2}{*}{Position of singularities, meV}} \\
No& \multicolumn{18}{|c|} {}  \\  \hline
1&4&6& 9 &15& 21& 24& 29& \multicolumn{2}{|c|}{35}& 41& 45& 52& 60& 66& \multicolumn{2}{|c|}{72}& 80& 87\\ \cline{9-10} \cline{16-17}
4&4& 6& 11.6& 18& 22& 25& 29& 32.5& 37& 42& 45& 54& 62& 65& 71& 75& 82& 87\\ \hline
\end{tabular}
\normalsize
\label{tab4}
\end{table}
In Refs. \cite{49} and \cite{50} devoted to tunneling in $Bi_{2}Sr_{2}CaCu_{2}O_8$, it was shown that singularities on the derivatives $dV/dI$ emerge at energies which are equal to the sum of the energy gap and one or two frequencies taken from experiments on Raman scattering. In our case, no data are available on the frequencies of lattice vibrations (or other elementary excitations) in SL. Hence it is not possible to associate individual peaks on the point contact spectra with interactions with any particular quasiparticles. Further investigations of SL with different sets of thicknesses of $PbTe-PbS$ layers must be carried out to find the nature of peaks.
\section{CONCLUSION}
On the basis of the entire body of investigations carried out by us, it can be concluded that SL are models of HTS in which all characteristic structural dimensions are enhanced by an order of magnitude, and the critical parameters are accordingly reduced, thus considerably simplifying their investigation. The following properties are common for SL and HTS.
\begin{enumerate}
\item Investigations of the fluctuational conductivity regions of SL, carried out by the standard resistive technique as well as by using the $R_{D}^{0}(T)$ and $R_{D}^{V}(T)$ dependences of point contacts, lead to the conclusion that depending on the electron mean free path and the binding force of the dislocation network, the initial pairing of electrons occurs either in the vicinity of the dislocation sites in the MD network plane, or as a result of interaction of electrons at the adjacent sites of neighboring dislocation networks. Apparently, an analogous situation is also characteristic of HTS materials in which the $CuO_2$ planes are the main elements responsible for superconductivity.

\item The use of point contacts led to the discovery of the metal-insulator transition for SL and HTS, starting in the region of SC fluctuations and terminating after a transition to the SC state. Investigations of the temperature dependence $R_{D}^{V}(T)$ for $Cu$-SL point contacts showed that the metal-insulator transition is apparently accompanied by a delocalization of carriers from the region of "potential wells" concentrated in the vicinity of structural elements responsible for the emergence of SC.

\item Both SL and HTS are characterized by large values of $2\Delta_{0}/kT_{c}\approx 10$. Anisotropic three-dimensional SL exhibit two gaps accompanied by a wide temperature interval of gapless superconductivity. The temperature dependences of the gaps cannot be described by the BCS theory. In the case of two gaps in an SL, the smaller gap at low temperatures is closer to the BCS value: $2\Delta_{1}(0)/kT_{c}\sim 3.5$. In quasi-two-dimensional SL or in HTS, the value of the gap observed in the fluctuational region of temperatures have a weak or no dependence on temperature. The gap emerges simultaneously with zero-dimensional fluctuations at $T\simeq 2T_{c}$ (for $YBaCuO$ superconducting fluctuations apparently emerge for $T\simeq 2.4~T_{c}\simeq 220~K$). For $T<T_c$, the temperature dependence of the energy gap for quasi-two-dimensional SL is nonmonotonic and associated with the localization of superconductivity on the MD network at low temperatures.

\item The $d^{2}F/dI^2$ curves for $Cu$-SL point contacts reveal peaks whose position on the energy scale is independent of temperature. Apparently, these peaks reflect the phonon structure of SL and as in the case of LSCO, their spectrum is restricted to an energy $\sim 90~meV$.

\item  The analogy in the properties of SL and HTS leads to the assumption that the superconductivity mechanism in SL is similar to that in HTS.

\end{enumerate}

This research was supported by the Scientific Council on HTS Problems and was carried out under projects No.35 "Mikrokontakt", No.364 "Astra", and No.308 "Stabil'nost'-77" of the State Program on "High-temperature Superconductivities".
\section{NOTATION}
Here $n_H$ is the Hall concentration of electrons, $\mu_{H}$ the electron mobility, $\sigma$  the conductivity, $\xi (T)$ the Ginsburg-Landau coherence length, $\tau$ the reduced temperature, \emph{l} the electron mean free path, $\Lambda$ the period of superlattice or $YBaCuO$ along the \emph{c}-axis, $\Delta$ the superconducting energy gap, $j_c$ the critical current density, $d$ the point contact diameter, $\rho$ the resistivity, and $R$ the resistance.


\begin{thebibliography}{}


\bibitem{1} V. Matijasevich and M.R. Beasley, in: Metal Superlattices. Artif. Struct. Mater., Amsterdam (1987).
\bibitem{2} O.A. Mironov, B.A. Savitskii, A.Yu. Sipatov, et al., \href{http://www.jetpletters.ac.ru/ps/184/article_3140.pdf}{Pis'ma Zh. Eksp. Teor. Fiz.} 48, 100 (1988) [\href{http://www.jetpletters.ac.ru/ps/1101/article_16651.pdf}{JETP Lett.} 48, 106 (1988)].
\bibitem{3} I.K. Yanson, N.L. Bobrov, L.F. Rybal'chenko, et al., \href{http://www.jetpletters.ac.ru/ps/225/article_3757.pdf}{Pis'ma Zh. Eksp. Teor. Fiz.} 49, 293 (1989) [\href{http://www.jetpletters.ac.ru/ps/1116/article_16905.pdf}{JETP Lett.} 49. 335 (1989)].
\bibitem{4} O.A. Mironov, S.V. Chistyakov, I.Yu. Skrylev, et al., \href{http://www.jetpletters.ac.ru/ps/255/article_4224.pdf}{Pis'ma Zh. Eksp. Teor. Fiz.} 50, 300 (1989) [\href{http://www.jetpletters.ac.ru/ps/1129/article_17115.pdf}{JETP Lett.} 50, 334 (1989)].
\bibitem{5} A.I. Fedorenko, B.A. Savitskij, A.Yu. Sipatov, et al., \href{http://web.kpi.kharkov.ua/krio/wp-content/uploads/sites/41/2013/07/APP-1900_new.pdf}{Acta Phys. Polon.} A77, 251 (1990).
\bibitem{6} T. Ishida, \href{http://iopscience.iop.org/article/10.1143/JJAP.28.L573/pdf}{Jap. J. Appl Phys.} 28, L573 (1989).
\bibitem{7} R.S.Liu, P.T. Wu, J.M. Liang, and L.J. Chen, \href{http://journals.aps.org/prb/pdf/10.1103/PhysRevB.39.2792}{Phys. Rev. B}39, 2792 (1989).
\bibitem{8} V.S. Fomenko, \href{http://www.twirpx.com/file/1578041/}{Emissive Properties of Materials} [in Russian], Naukova Dumka, Kiev (1981).
\bibitem{9} V.G. Kantser and N.M. Malkova, in: Abstracts of Papers to XXVI All-Union Conf. on Low Temp. Phys., Donetsk (1990).
\bibitem{10} V.I. Kaidanov and Yu. I. Ravich, \href{http://ufn.ru/ufn85/ufn85_1/Russian/r851b.pdf}{Usp. Fiz. Nauk} 145, 51 (1985) [Sov. Phys. Usp. 28, 31 (1985)].
\bibitem{11} A.I. Golovashkin, Î.M. Ivanenko, and Ê. V. Mitsen, Sverkhprovodimost': Fiz., Khim., Tekh. 2, 82 (1989).
\bibitem{12} V.P. Galaiko, \href{http://fntr.ilt.kharkov.ua/fnt/pdf/13/13-10/f13-1102r.pdf}{Fiz. Nizk. Temp.} 13, 1102 (1987) [Sov. J. Low Temp.Phys. 13, 627 (1987)].
\bibitem{13} Anil Khurana, \href{http://scitation.aip.org/content/aip/magazine/physicstoday/article/42/4/10.1063/1.2810965}{Phys. Today} 42, 17 (1989).
\bibitem{14} S.I. Zolotov, A.N. Kovalev, V.I. Paramonov, and À.E. Yunovich, Fiz. Tekh. Poluprovodn. 19, 616 (1985) [Sov. Phys. Semicond. 19, 382 (1985)].
\bibitem{15} S.S. Borisova, I.F. Mikhailov, and L.P. Shpakovskaya, Kristallografiya 31, 651 (1986) [Sov. Phys. Crystallogr. 31, 384 (1986)].
\bibitem{16} M. Gol'tsman, Z.M. Dashevskii, V.I. Kaidanov, and N.V. Kolomoets, \href{http://libarch.nmu.org.ua/handle/GenofondUA/65326}{Thin Film Thermal Elements}: Physics and Applications [in Russian], Nauka, Moscow (1985).
\bibitem{17} Parker E.H.  and D. Williams, \href{http://www.sciencedirect.com/science/article/pii/0040609076902030}{Thin Solid Films} 35, 373 (1976).
\bibitem{18} L.G. Aslamazov and A.I. Larkin, Fiz. Tverd. Tela (Leningrad) 10, 1104 (1968) [Sov. Phys. Solid State 10, 875 (1968)]
\bibitem{19} K. Maki, \href{http://ptp.oxfordjournals.org/content/40/2/193.full.pdf+html}{Progr. Theor. Fiz.} 40, 193 (1968).
\bibitem{20} R.S. Thompson, \href{http://journals.aps.org/prb/abstract/10.1103/PhysRevB.1.327}{Phys. Rev. B}1, 327 (1970).
\bibitem{21} K. Maki and R.S. Thompson, \href{https://journals.aps.org/prb/abstract/10.1103/PhysRevB.39.2767}{Phys. Rev. B}39, 2767 (1989).
\bibitem{22} Y. Matsuda, T. Hirai, and S. Komiyama, \href{http://www.sciencedirect.com/science/article/pii/0038109888902542}{Solid State Commun.} 68, 103 (1988).
\bibitem{23} I.O. Kulik, \href{http://fntr.ilt.kharkov.ua/fnt/pdf/14/14-2/f14-0209r.pdf}{Fiz. Nizk. Temp.} 14, 209 (1988) [Sov. J. Low Temp. Phys. 14, 116 (1988)].
\bibitem{24} M.W. Shafer, T. Penney, B.L. Olson, et al., \href{http://journals.aps.org/prb/abstract/10.1103/PhysRevB.39.2914}{Phys. Rev. B}39, 2914 (1989).
\bibitem{25} I.O. Kulik, \href{http://www.sciencedirect.com/science/article/pii/0378436384901761}{Physica B} 126, 280 (1984).
\bibitem{26} S.N. Burmistrov and L.B. Dubovskii, \href{http://www.sciencedirect.com/science/article/pii/037596018990827X}{Phys. Lett.} 136, 332 (1989).
\bibitem{27} O.A. Mironov, S.V. Chistyakov, I.Yu. Skrylev, et al., in: Abstracts of Papers to the 3rd All-Union Symp. on Nonhomogeneous Electron States, Novosibirsk (1989).
\bibitem{28} N.E. Alekseevskii, A.V. Mitin, V.I. Nizhankovskii, et al., Sverkhprovodimost': Fiz., Khim., Tekh. 2, 40 (1989).
\bibitem{29} N. L.Bobrov, L.F. Rybal'chenko, M.A. Obolenskii, and V.V. Fisun, \href{http://fntr.ilt.kharkov.ua/fnt/pdf/11/11-9/f11-0925r.pdf}{Fiz. Nizk. Temp.} 11, 925 (1985) [Sov. J. Low Temp. Phys. 11, 510 (1985)],  \href{https://arxiv.org/pdf/1603.02598.pdf}{arXiv:1603.02598}.
\bibitem{30}  B. N. Engel, G. G. Ihas, E. D. Adams and C. Fombarlet, Prib. Nauchn. Issled., No. 12, 132 (1984)[\href{http://scitation.aip.org/content/aip/journal/rsi/55/9/10.1063/1.1137965}{Review of Scientific Instruments}, 55, 1489 (1984)].
\bibitem{31} O.I. Shklyarevskii, N.N. Gribov, and Yu.G. Naidyuk, \href{http://fntr.ilt.kharkov.ua/fnt/pdf/9/9-10/f09-1068r.pdf}{Fiz. Nizk.Temp.} 9, 1068 (1983) [Sov. J. Low Temp. Phys. 9, 553 (1983)].
\bibitem{32} O.I. Shklyarevskii, A.M. Duif, A.G. M. Jansen, and P. Wyder, \href{http://journals.aps.org/prb/abstract/10.1103/PhysRevB.34.1956}{Phys. Rev. B}34, 1956 (1986).
\bibitem{33} O.I. Shklyarevskii, I.K. Yanson, and N.N. Gribov, Fiz. Nizk. Temp. 14, 479 (1988) [Sov. J. Low Temp. Phys. 14, 263 (1988)].
\bibitem{34} I.K. Yanson, L.F. Rybal'chenko, V.V. Fisun, et al., \href{http://fntr.ilt.kharkov.ua/fnt/pdf/14/14-11/f14-1157r.pdf}{Fiz. Nizk. Temp.} 14, 1157 (1988) [Sov. J. Low Temp. Phys. 14, 639 (1988)], \href{https://arxiv.org/pdf/1512.06416.pdf}{arXiv:1512.06416}.
\bibitem{35} L.F. Rybal'chenko, I.K. Yanson, V.V. Fisun, et al., \href{http://fntr.ilt.kharkov.ua/fnt/pdf/16/16-8/f16-1033r.pdf}{Fiz. Nizk. Temp.} 16, 1033 (1990) [Sov. J. Low Temp. Phys. 16, 602 (1990)].
\bibitem{36} I.K. Yanson, L.F. Rybal'chenko, V.V. Fisun, et al., \href{http://fntr.ilt.kharkov.ua/fnt/pdf/15/15-8/f15-0803r.pdf}{Fiz. Nizk. Temp.} 15, 803 (1989) [Sov. J. Low Temp. Phys. 15, 445 (1989)].
\bibitem{37} I.K. Yanson, V.V. Fisun, N.L. Bobrov, and L.F. Rybal'chenko, \href{http://www.jetpletters.ac.ru/ps/141/article_2443.pdf}{Pis'ma Zh. Eksp. Teor. Fiz.} 45, 425 (1987) [\href{http://www.jetpletters.ac.ru/ps/1244/article_18813.pdf}{JETP Lett.} 45, 543 (19-87)].
\bibitem{38} I.K. Yanson, L.F. Rybal'chenko, V.V. Fisun, et al., \href{http://fntr.ilt.kharkov.ua/fnt/pdf/14/14-8/f14-0886r.pdf}{Fiz. Nizk. Temp.} 14, 886 (1988) [Sov. J. Low Temp. Phys. 14, 487 (1988)].
\bibitem{39} L.F. Rybal'chenko, I.K. Yanson, N.L. Bobrov, et al., \href{http://fntr.ilt.kharkov.ua/fnt/pdf/16/16-1/f16-0058r.pdf}{Fiz. Nizk. Temp.} 16, 58 (1990) [Sov J. Low Temp. Phys. 16, 30 (1990)].
\bibitem{40} J.J. Gniewek, J.C. Moulder, and R.H. Kropschot, in: Proc. Tenth Int. Conf. on Low Temp. Phys., Moscow (1967).
\bibitem{41} I.Ê. Yanson, L.F. Rybal'chenko, N.L. Bobrov, and V.V. Fisun, \href{http://fntr.ilt.kharkov.ua/fnt/pdf/13/13-8/f13-0873r.pdf}{Fiz. Nizk. Temp.} 13, 873 (1987) [Sov. J. Low Temp. Phys. 13, 501 (1987)].
\bibitem{42} A.V. Zaitsev, Zh. Eksp. Teor. Fiz. 86, 1742 (1984) [\href{http://www.jetp.ac.ru/cgi-bin/dn/e_059_05_1015.pdf}{Sov. Phys. JETP} 59, 1015 (1984)].
\bibitem{43} O. Iwanyshyn and J.T. Smith, \href{http://journals.aps.org/prb/abstract/10.1103/PhysRevB.6.120}{Phys. Rev. B}6, 120 (1972).
\bibitem{44} Gerald B. Arnold, \href{http://link.springer.com/article/10.1007/BF00681510}{J. Low Temp. Phys.} 59, 143 (1985).
\bibitem{45} P.S. van Son, M. van Kempen, and P. Wyder, \href{http://journals.aps.org/prl/abstract/10.1103/PhysRevLett.59.2226}{Phys. Rev. Lett.} 59, 2226 (1987).
\bibitem{46} J. Geerk, X.X. Xi, and G. Linker. \href{http://link.springer.com/article/10.1007/BF01314271}{Z. Phys.} B73, 329 (1988).
\bibitem{47} V.Z. Kresin, \href{http://www.sciencedirect.com/science/article/pii/0038109887901190}{Solid State Commun.} 63, 725 (1987).
\bibitem{48} H. Ikuta, A. Maeda, K. Uchinokura, and S. Tanaka, \href{http://iopscience.iop.org/article/10.1143/JJAP.27.L1038/meta}{Jpn. J. AppL Phys.} 27, L1038 (1989).
\bibitem{49} D. Shimada, N. Miyakawa, T. Kido, and N. Tsuda, \href{http://journals.jps.jp/doi/10.1143/JPSJ.58.387}{J. Phys. Soc. Jpn.} 58, 387 (1989).
\bibitem{50} N. Miyakawa, D. Shimada, T. Kido, and N. Tsuda, \href{http://journals.jps.jp/doi/10.1143/JPSJ.58.1141}{J. Phys. Soc. Jpn.} 58, 1141 (1989).
\bibitem{51} S.S. Torardi, E.M. McCarron, M.A. Subramanian, et al., in: High-Temperature Superconductors, D. Nelson, M. Wittingham, and T. George (eds.) [Russian translation], Mir, Moscow (1988).

\end{thebibliography}
\end{document}